\documentclass[%
 aip,
 amsmath,amssymb,
reprint,
floatfix]{revtex4-1}

\usepackage{graphicx}% Include figure files
\usepackage{dcolumn}% Align table columns on decimal point
\usepackage{bm}% bold math
\usepackage[utf8]{inputenc}
\usepackage[T1]{fontenc}
\usepackage{mathptmx}
\usepackage{etoolbox}

\usepackage{lipsum}
\usepackage{siunitx}
\DeclareMathOperator{\Tr}{Tr}
\newcommand\Tstrut{\rule{0pt}{2.6ex}}         % = `top' strut

\usepackage{hyperref}
\usepackage{xcolor}
\hypersetup{
    colorlinks,
    linkcolor={red!50!black},
    citecolor={blue!50!black},
    urlcolor={blue!80!black}
}
\usepackage{ragged2e}
\usepackage[caption=false]{subfig}

\usepackage{float}
\usepackage[normalem]{ulem}

\newcommand{\CM}{cm$^{-1}$}

\makeatletter
\def\@email#1#2{%
 \endgroup
 \patchcmd{\titleblock@produce}
  {\frontmatter@RRAPformat}
  {\frontmatter@RRAPformat{\produce@RRAP{*#1\href{mailto:#2}{#2}}}\frontmatter@RRAPformat}
  {}{}
}%
\makeatother
\begin{document}

\preprint{}

\title{Prediction of Fluorescence Quantum Yields using the Extended Thawed Gaussian Approximation}
% Force line breaks with \\

\author{Michael Wenzel}
%  \email{michael.wenzel@uni-wuerzburg.de}
%  \altaffiliation[Also at ]{Institut für Physikalische und Theoretische Chemie, Universität Würzburg.}%Lines break automatically or can be forced with \\
 \affiliation{Institut für Physikalische und Theoretische Chemie, Universität Würzburg, Emil-Fischer Str. 42, 97074 Würzburg, Germany%\\This line break forced with \textbackslash\textbackslash
}%

\author{Roland Mitric$^*$}%
\email{roland.mitric@uni-wuerzburg.de}
\affiliation{Institut für Physikalische und Theoretische Chemie, Universität Würzburg, Emil-Fischer Str. 42, 97074 Würzburg, Germany%\\This line break forced with \textbackslash\textbackslash
}%

\date{\today}% It is always \today, today,

\begin{abstract}
Spontaneous emission and internal conversion rates are calculated within harmonic approximations and compared to results obtained within the semi-classical extended thawed Gaussian approximation. This is the first application of the ETGA in the calculation of internal conversion and emission rates for real molecular systems, namely   formaldehyde, fluorobenzene, azulene and a dicyano-squaraine dye. The viability of the models as black-box tools for prediction of spontaneous emission and internal conversion rates is assessed. All calculations were done using a consistent protocol in order to investigate how different methods perform without previous experimental knowledge, using DFT/TD-DFT with the B3LYP, PBE0, $\omega$B97XD and CAM-B3LYP functionals. Contrasting the results with experimental data shows that there are further improvements required before theoretical predictions of emission and internal conversion rates can be used as reliable indicator for the photo-luminescence properties of molecules. We find that the extended thawed Gaussian approximation performs rather similar to the vertical harmonical model. Including anharmonicities in the calculation of internal conversion rates has a moderate effect on the quantitative results in the studied systems. The emission rates are fairly stable with respect to computational parameters, but the internal conversion rate reveals itself to be highly dependent on the choice of the spectral lineshape function, particularly the width of the Lorentzian function, associated with homogeneous broadening.
\end{abstract}

\maketitle

% \section{\label{sec:level1}First-level heading:\protect\\ The line
% break was forced \lowercase{via} \textbackslash\textbackslash}

\section{\label{sec:level1}Introduction}

The exchange of energy between electronic and nuclear degrees of freedom is the driving force of nonadiabatic processes. In the case of excited molecules, this exchange is known as internal conversion, and constitutes a competing channel to spontaneous emission, where energy is transferred to the modes of the electromagnetic field, detectable as fluorescence or, if the transition involves states of different spin, phosphorescence. Highly efficient organic photoluminescent compounds are desirable, e.g. for organic light emitting devices but also as markers and sensors in bio-imaging.\cite{Frangioni2003,Kim2003,Kuang2003,Peng2011,Zhang2018,Hudson2021,Song2022}   The discovery and design of suitable compounds with strong fluorescence and high quantum yields thus continues to be of interest.\cite{Samuel2007,Zhu2013,Shuai2016,Hong2021}

Excited electronic states can relax by different mechanisms. Ultrafast relaxation is known to be mediated by energetically accessible conical intersections.\cite{Domcke2004,Polli2010,Fuji2010} In such cases, the initially excited state is usually short lived and fluorescence from this state is not observed. A well known process that falls into this regime is for example the photoisomerization of azobenzene, that takes place on a femtoseconds time scale and is of interest due to its possible applications in photoswitches.\cite{HARTLEY1937,Wachtveitl1997,Tamai2000,Schultz2003,Beharry2011} The modeling of the intricate nonadiabatic dynamics that involve nuclei and electrons alike, requires methods that go beyond the Born-Oppenheimer approximation. For systems with a limited amount of degrees of freedom, a fully quantum mechanical treatment may be used, like multilayer multiconfigurational time-dependent Hartree (ML-MCTDH)\cite{Wang2003,Manthe2008,Liu2018,Green2021} or time-dependent density matrix renormalization group method (TD-DMRG).\cite{Wang2021,Ren2021}%\cite{Green2021,Wang2021,Ren2021} 
These approaches can be computationally demanding, and treatment of large systems remains challenging. Mixed quantum-classical methods, which utilize classical trajectories as guiding principle, remain affordable when a full quantum mechanical treatment is no longer possible, schemes such as fewest-switches surface hopping\cite{Tully1990,Mitric2009,Barbatti2011,Roehr2013,Hoche2017,Malhado2014,Cui2014,Xie2019} (FSSH) algorithms, multi-configurational Ehrenfest dynamics or mixed approaches.\cite{Prezhdo1997,C.Tully1998,Drukker1999,Hack2000,Li2005,Subotnik2010,Shalashilin2011,Vacher2014,CrespoOtero2018,Schuurman2018,Ma2018,Danilov2023}

%For systems with a limited amount of degrees of freedom, a fully quantum mechanical treatment may be used, like multilayer multiconfigurational time-dependent Hartree (ML-MCTDH) or time-dependent density matrix renormalization group method (TD-DMRG)
%In the smallest systems one may attempt direct integration of the time-dependent Schrödinger equation\cite{Kosloff1988} and for systems with more than just a few degrees of freedom, a fully quantum mechanical treatment remains possible, using multilayer multiconfigurational time-dependent Hartree (ML-MCTDH)\cite{Wang2003,Manthe2008,Liu2018} or time-dependent density matrix renormalization group method (TD-DMRG).\cite{Green2021,Wang2021,Ren2021}

Not all photo-induced reactions are ultra fast. Access to a conical intersection can be hindered by an energetic barrier on the same potential energy surface and the rate determining step is then given by crossing of this barrier.\cite{Hudock2007,Blacker2013,Hoche2019,Kramers1940,Kohn2019,Lin2020} The direct simulation of the dynamics may become demanding, as long propagation times would be required to observe the reaction of interest, although FSSH with decoherence correction is capable of modeling such processes.\cite{Landry2012,Jain2015} Classical transition state theory,\cite{Rice1927,Kassel1928,Eyring1931,Eyring1935,Evans1935,Laidler1983} or ring-polymer molecular dynamics rate theory, which takes quantum effects such as tunneling into account, are viable options in such cases.\cite{Tromp1987,Richardson2009,Habershon2013}Transition state theory (TST) is not limited to adiabatic processes\cite{Marcus1964} and the development of nonadiabatic TST that include quantum effects is  progressing and constitute a valuable part in the modeling of chemical dynamics.\cite{Zahr1975,Miller1979,Heller1983,Lorquet1988,Richardson2018,Ansari2022,Trenins2022} 
%But the main contribution to rate theories is typically centered on a semiclassical path through a minimum energy crossing point of diabatic potentials.  

Another option to treat processes that involve tunneling of nuclei is time-dependent perturbation theory in the form of Fermi's golden rule (FGR),\cite{Dirac1927,Fermi1950} as long as the transition takes place between weakly coupled electronic states. This is a fair assumption for slow internal conversion, that is not dominated by quick relaxation through a conical intersection. The viability and the conditions for this description of intramolecular nonradiative relaxation have been discussed in great detail in literature, leading also to the established energy gap law for radiationless transitions in large molecules and efficient rate expressions for internal 
%Zwischen valiev 2020 und 2019 muss noch ein Zitat rein, ansonsten muss alles neu nummeriert werden.
conversion.\cite{Lin2004,Siebrand1967,Bixon1968,Englman1970,Freed1970,Nitzan1973,Nitzan1973a,Fujimura1975,Plotnikov1979,Valiev2018,Valiev2019,Valiev2020,Valiev2021,Erker2022}
FGR is also applicable to the prediction of spontaneous emission rates, and provides a common framework to model radiative and nonradiative processes at the same theoretical level. But a direct evaluation of the rate expression within the formalism is usually not possible, as the exact eigenstates of the initial and final potential are not known. This problem can be alleviated by invoking the harmonic approximation for all involved potentials. Required matrix elements have then an analytic solution,\cite{Hazra2003,Hazra2004,Dierksen2005,Niu2008,Barone2009,Bloino2010,Ferrer2012,Sorour2022} likewise the equivalent correlation function of a time-dependent formulation of the problem which avoids an explicit summation over states.\cite{Hayashi1998,Peng2010,Peng2013,Banerjee2016,Miyazaki2022}

The approximation works well for transitions to the vibrational ground state or excited vibrational states with only a few quanta of energy. This is typically the case for absorption or spontaneous emission, but internal conversion can involve highly excited vibrational states, as energy conservation prohibits the transition to low lying states when there is an appreciable energy gap between the involved electronic states. The anharmonicity of the potential may no longer be negligible and the validity of purely harmonic models is no longer evident.\cite{Ianconescu2011}  
Semi-classical methods for wave packet propagation are not limited to globally harmonic potentials\cite{lasser_lubich_2020, Heller1975,Heller1976,Heller1981,Heller1981a,Hagedorn1980,Hagedorn1985,Hagedorn1998,Herman_1984,Kluk1986,Walton1996} and provide a pathway to methods that are a good compromise of efficiency and accuracy. Highly accurate results can be obtained even for potentials with pronounced anharmonicity and initial states of any shape, using swarms of Gaussian wave packets in a semi-classical initial value representation approach.\cite{Ianconescu2011,Ianconescu2013} But achieving convergence of the results with respect to number of trajectories can be expensive, in particular in large systems and under consideration of all degrees of freedom. 
This work focuses thus on a single-trajectory based procedure from a larger family of Gaussian wavepacket dynamics,\cite{J.L.Vanicek2023} the semi-classical Extended Thawed Gaussian approximation (ETGA).\cite{Patoz2018,BEGUSIC2018_3TGA,Begusic2019,Prlj2020,Begusic2020,Begusic2022} Its efficiency enables a treatment of molecules with all internal degrees of freedom. The working equations for internal conversion are presented and it is investigated whether the method's semi-classical nature constitutes a substantial improvement over the adiabatic harmonic model (AH) and the vertical harmonic model (VH), purely harmonic approaches, which are used as reference methods.\cite{Hazra2003,Hazra2004,Ferrer2012,Peng2013,Banerjee2016} These reference methods differ in the expansion point for the harmonic approximation of the electronic potential, the adiabatic model uses the optimized geometry of the PES while the vertical model uses the geometry corresponding to a vertical transition. The working equations of these methods have been reported in references \onlinecite{Tang2003,Ianconescu2004,Peng2007,Peng2010,Borrelli2011,Baiardi2013,Borrelli2015,Banerjee2016,Cerezo2023} and are not repeated herein. The goal of this work is to assess the robustness and quality of the ETGA for the calculation of internal conversion and spontaneous emission rates at the FGR level, applied to molecular systems with all internal degrees of freedom included. This presents the next logical step after tests of the method in one-dimensional model systems in a previous publication.\cite{Wenzel2023}

 \begin{figure}
  \centering
  \includegraphics[width=.33\textwidth, trim=0 0.0cm 0 0.0cm, clip]{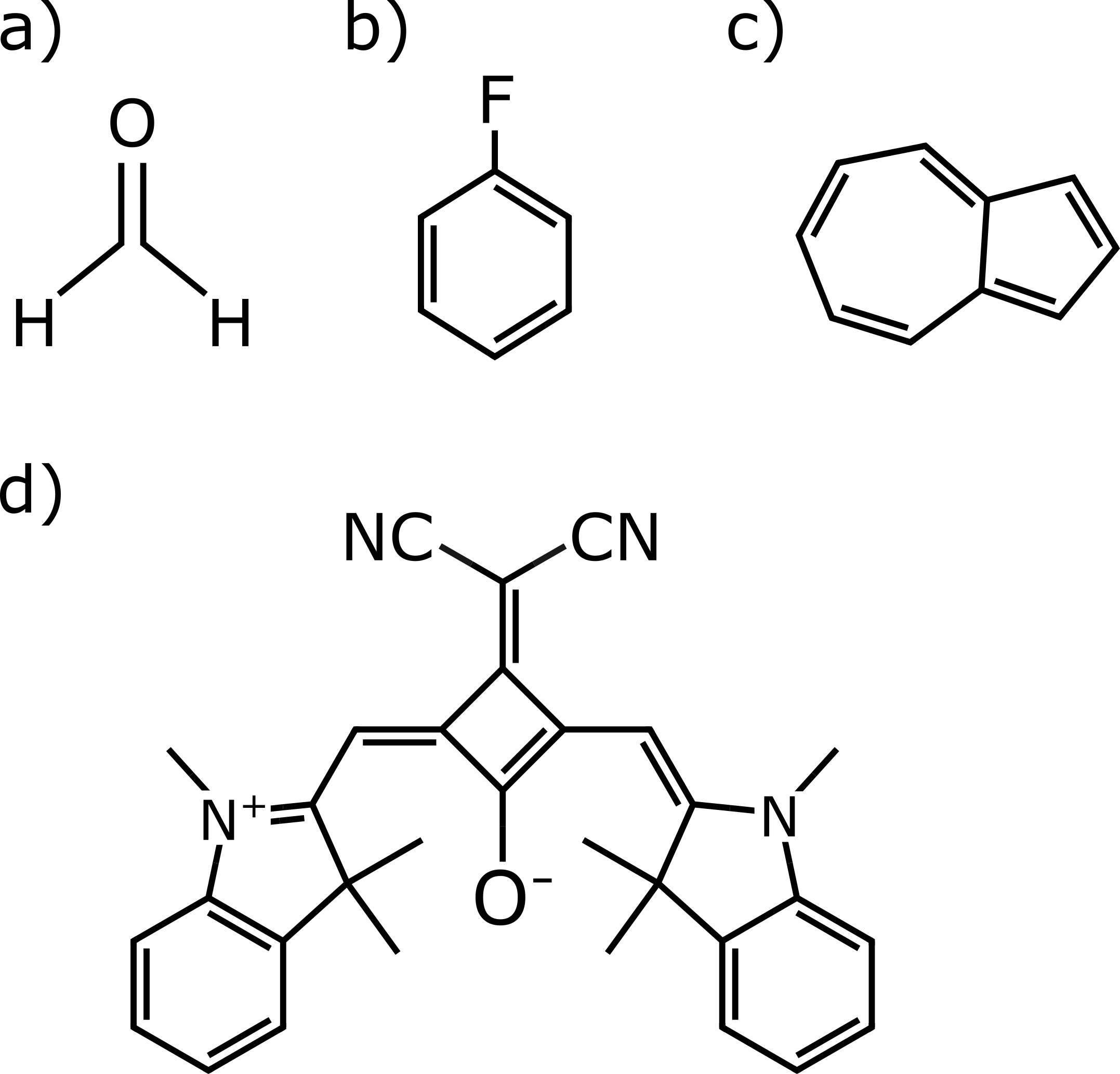}
  \caption{Molecular structures of a) formaldehyde, b) fluorobenzene, c) azulene and d) squaraine dye.}
  \label{fig:molecular_structures}
\end{figure}

The results are presented for the $S_1\rightarrow S_0$ transition of formaldehyde, fluorobenzene and a squaraine dye. For azulene, a well known exception to Kasha's rule and also part of the test set, the fluorescence rate is calculated for the $S_2\rightarrow S_0$ transition, while the competing internal conversion rate is assumed to be dominated by $S_2\rightarrow S_1$ internal conversion. The molecular structures of the four test systems are given in Fig. \ref{fig:molecular_structures}.

\section{Theory}

\subsection{Spontaneous Emission and Internal Conversion Rates}

The rotational averaged spontaneous emission spectrum within first order time-dependent perturbation theory in form of an  auto-correlation function\cite{Tannor2006,Niu2010,Begusic2019} is given by
\begin{align}
\sigma_\mathrm{SE}(\omega) &= \frac{\omega^3}{6\hbar c^3\pi^2\varepsilon_0} \int^\infty_{-\infty} dt \exp(-i\omega t) \langle \phi_{in}|U_i^\dagger \vec \mu^\dagger_{fi} U_f \vec \mu_{fi} |\phi_{in}\rangle \label{eq:emission_spectrum},
\end{align}
with
\begin{align}
\vec \mu_{fi}&=\left( \begin{matrix}
    \mu_{fi,x} &= \langle \psi_f|\mu_x|\psi_i\rangle_r \\
    \mu_{fi,y} &= \langle \psi_f|\mu_y|\psi_i\rangle_r \\
    \mu_{fi,z} &= \langle \psi_f|\mu_z|\psi_i\rangle_r \\
\end{matrix}\right),
\end{align}
such that
\begin{align}
    \langle \phi_{in}|U_i^\dagger \vec \mu^\dagger_{fi} U_f \vec \mu_{fi} |\phi_{in}\rangle &\equiv \sum_{\alpha \in x,y,z} \langle \phi_{in}|U_i^\dagger \mu^\dagger_{fi,\alpha} U_f  \mu_{fi,\alpha} |\phi_{in}\rangle.
\end{align} 
We have assumed that the system is initially in the vibronic state $|\phi_{in}\rangle$, specifying the adiabatic electronic state $|\psi_i\rangle$ with first index $i$ and the associated vibrational state with index $n$. The propagator $U_i$ determines the time evolution of a nuclear wave packet on the potential energy surface of electronic state $i$, while the matrix elements of the dipole operator with respect to the electronic adiabatic state basis are given by $\vec \mu_{fi}$, obtained by integration of the electronic coordinates $r$. 

The rate of spontaneous emission originating from the initial state to the vibronic manifold $\{ \phi_{fm} \}$ associated with electronic state $|\psi_f\rangle$ is then given by the integral of the emission spectrum,

\begin{equation}
    k_\mathrm{SE} = \int_0^\infty d\omega \ \sigma_\mathrm{SE}(\omega).
\end{equation}

Internal conversion is treated within the same perturbative framework, to yield the following expression\cite{Ianconescu2011,Ianconescu2013,Humeniuk2020} for the nonradiative rate,
\begin{equation}
    k_\mathrm{IC} = \frac{1}{\hbar^2}\int ^\infty_{-\infty} dt \langle \phi_{in}|U^\dagger_i T^\dagger_{fi} U_f T_{fi}|\phi_{in} \rangle,
    \label{eq:k_ic}
\end{equation}
with nuclear kinetic energy operator $T$ as driving force of the nonadiabatic transition. A transition matrix element in the basis of the adiabatic electronic states and expressed in mass weighted nuclear normal coordinates $q$ of the initial electronic state, is taken to be
\begin{equation}
 T_{fi} =- \tau_{fi}^T\nabla_q,\label{eq:nac}
\end{equation}
including only the first order nonadiabatic coupling element $\tau_{fi}=\hbar^2\langle \psi_f| \nabla_q | \psi_i \rangle$, whereas second order derivative contributions are neglected. The law of energy conservation requires that internal conversion takes place between states with the same energy but for the analysis of internal conversion rate methods it is useful to define the internal conversion rate spectrum
\begin{equation}
    k_\mathrm{IC}(\omega) = \frac{1}{\hbar^2}\int ^\infty_{-\infty} dt \langle \phi_{in}|U^\dagger_i T^\dagger_{fi} U_f T_{fi}|\phi_{in} \rangle\exp(i\omega t),
    \label{eq:kic_spectrum}
\end{equation}
a quantity that shows how the internal conversion rate would change if the energy gap of the initial and final electronic states is varied, while all other parameters are kept the same. Evaluating this spectrum at $\hbar \omega= 0 \ \mathrm{eV}$ yields the internal conversion rate. The effect of a decrease of the energy gap between final and initial electronic states on the internal conversion rate becomes clear by following the function values along the negative abscissa, while the effect of an increase of the energy gap corresponds to the spectrum at positive values of $\hbar\omega$.

With the internal conversion and spontaneous emission rates available, the quantum yield of spontaneous emission can be calculated according to
\begin{equation}
    \Phi_\text{QY} = \frac{k_\mathrm{SE}} {k_\mathrm{SE} + k_\mathrm{IC}}.
\end{equation}

Equations \ref{eq:emission_spectrum} and \ref{eq:k_ic} correspond to transitions from a single vibronic state $|\phi_{in}\rangle$. This is a suitable model for  emission processes that occur in isolated molecules, like a transition from a vibrational ground state of a S$_1$ state to the S$_0$ manifold.

Equations for a canonical ensemble in thermal equilibrium at temperature $T$ are obtained by summing over a Boltzmann factor weighted set of initial states. It is assumed that the temperature is low enough that only vibrational states of the same initial electronic state need to be considered. In this case it suffices to sum  over index $n$ of the initial vibronic state. 

The specific coupling operators are replaced  with a generic operator $K$ in the following, to facilitate a general derivation and to focus solely on the correlation function, which is at the core of the time-dependent formulation. The correlation function for an isolated molecule is
\begin{equation}
    C(t) = \langle \phi_{in}|U_i^\dagger K^\dagger_{fi} U_f K_{fi}|\phi_{in}\rangle \label{eq:correlation_function_k},
\end{equation}
and the canonical ensemble correlation function is obtained as
\begin{align}
    C(t,T) &= \frac{1}{Z_i}\sum_n \exp(-\beta E_{in})C(t) \\
    &= \Tr [\rho_i U^\dagger_i K^\dagger_{fi} U_f K_{fi}] \label{eq:canonical_correlation_function},
\end{align}
with the canonical density operator ${\rho_i = \sum_n \frac{\exp(-\beta E_{ni})}{Z_i} |\phi_{in}\rangle \langle \phi_{in} |}$, the partition function $Z_i$ of the vibronic manifold associated with the initial electronic state and $\beta = 1/(k_B T)$. Replacing the isolated-case correlation function with its thermal pendant enables the treatment of systems in thermal equilibrium.

The canonical correlation function can be expressed in a form that is amenable to wave packet propagation methods, such as the Thawed Gaussian Approximation (TGA), which is introduced thereafter. A short derivation of this formulation is given in the following, limited to a few expositive steps, since a comprehensive treatment was already given recently in Ref. \onlinecite{Begusic2020}, covering these steps in detail.

The following relations 
\begin{align}
    \rho_i = \rho_i^{1/2}\rho_i^{1/2}, \ [\rho_i,H_i] = 0
\end{align}
and the cyclic property of the trace allow us to rearrange Eq.~\ref{eq:canonical_correlation_function} to
\begin{equation}
    C(t,T) = \Tr [\rho_i^{1/2}K^\dagger_{fi} U_f K_{fi} U_i^\dagger \rho^{1/2}_i ].
\end{equation}
Taking the trace with respect to nuclear normal coordinates $\{q'\}$ and inserting a completeness relation $1=\int dq |q\rangle \langle q|$, yields
\begin{equation}
    C(t,T) = \int \int dq' dq \langle q'| \rho_i^{1/2}K^\dagger_{fi}|q\rangle \langle q| U_f K_{fi}  \rho^{1/2}_i U_i^\dagger|q'\rangle.
\end{equation}
With the following definitions
\begin{align}
    \phi_0(q,q') &= \langle q| K_{fi} \rho_i^{1/2}|q'\rangle \label{eq:thermal_phi_0}\\
    \phi_t(q,q') &= \exp(-i ( H_f(q)-H_i(q'))t/\hbar ) \phi_0(q,q')\\
    U(q,q',t) &= \exp(-i ( H_f(q)-H_i(q'))t/\hbar ) \\
    \bar q &= (q,q'),
\end{align}
the expression for the canonical  correlation function becomes
\begin{align}
    C(t,T) &= \int d\bar q \ \phi_0^*(\bar q)U(\bar q,t)\phi_0(\bar q)\\
     &= \int d\bar q \ \phi_0^*(\bar q)\phi_t(\bar q).
\end{align}
The calculation of the canonical ensemble correlation function is now recognizable as wave packet autocorrelation, requiring the propagation of a $2N$ dimensional wave packet satisfying the following time-dependent Schrödinger equation,
\begin{align}
    i\hbar \dot \phi_t(\bar q) &= \bar H(\bar q )\phi_t(\bar q) \\
    \bar H(\bar q) &= H_f(q) - H_i(q'). \label{eq:thermal_vibronic_hamiltonian}
\end{align}

This means that methods for solving the time-dependent Schrödinger equation or equivalently methods for the propagation of a wave packet, can be applied the thermal problem in the same manner as in the isolated case. 

\subsection{Thawed Gaussian Approximation}

The TGA can be viewed as semi-classical extension of the harmonic approximation, based on the propagation of a thawed Gaussian wave packet on a time-dependent potential that is obtained by harmonic approximation of the PES at the current center of the wave packet.\cite{Heller1975,Heller1976} The isolated molecule case is treated first and then transferred to the thermal problem.

There are two essential parts to the method, for one the Ansatz for time-dependent wave packet in form of a thawed Gaussian,

\begin{equation}
    \phi_t(q) = \exp \left(\frac{i}{\hbar} \left[\frac{1}{2}(q-q_t)^TA_t(q-q_t)+p^T_t(q-q_t) + \gamma_t \right] \right)\, ,\label{eq:tga_definition}
\end{equation}

and the replacement of the exact potential of the final state by a time-dependent harmonic approximation that is obtained by expanding the true PES harmonically about a classical trajectory that lies at the center of the time-dependent wave packet as it propagates to obtain the local harmonic approximation (LHA) for the potential,

\begin{equation}
V_t(q) = V(q_t) + V'_t (q-q_t) + \frac{1}{2}(q-q_t)^T V''_t(q-q_t)\,\label{eq:lha_potential}.
\end{equation}

Inserting the Ansatz and potential into the time-dependent Schrödinger equation yields a set of differential equations for the  parameters that encode the time dependency of the wave packet,
\begin{align}
    \dot q_t &= p_t, \quad \dot p_t = -V'_t \label{eq:eom_qp_t}, \\
    \dot A_t &= -A_t A_t - V''_t\label{eq:eom_A_t}, \\
    \dot \gamma_t &= \frac{i\hbar}{2}\Tr(A_t) + L_t\label{eq:eom_gamma_t}.
\end{align}

The parameters $(q_t,p_t)$ follow Hamilton's equations for a classical trajectory. Parameter $A_t$ regulates the width and enables the Gaussian wave packet to contract or spread in response to the potential as it propagates. Normalization of the wave packet is ensured by the last parameter $\gamma_t$, and leads to a phase factor based on the classical Lagrangian $L_t=\frac{p_t^2}{2} - V(q_t)$, evaluated along the trajectory.
The solution of this set of differential equations is equivalent to a wave packet propagation based on a time evolution operator constructed with the time-dependent LHA potential,
\begin{align}
    U_t &= \mathcal{T} \exp(-\frac{i}{\hbar} \int_0^t H_{t'} dt'),\\
    U_t\phi_0 &= \phi_t
\end{align}
where $\mathcal T$ takes care of the time-ordering in the integral and a vibronic Hamiltonian given by
\begin{equation}
H_t = -\frac{\hbar^2}{2}\nabla_q^2 + V_t(q).
\end{equation}

\subsection{Extended Thawed Gaussian Approximation}
Up to this point we have assumed that our initial state has the form of a Gaussian function, without consideration for the coupling operator sandwiched between the initial state and the propagator. The transition dipole operator limited to Franck-Condon and Herzberg-Teller terms\cite{Franck1926,Condon1928,Herzberg1933} and the nonadiabatic coupling operator as defined in Eq. \ref{eq:nac}, that contains a derivative that acts on the nuclear wave packet, give rise to an initial state of the following form 
\begin{equation}
    K\phi_0  = c_0 + c_1^T(q-q_0)\phi_0, \label{eq:K_acting_on_phi}
\end{equation}
where $c_0$ is a scalar quantity while $c_1$ is vector with the same dimension as $q$.
This leads to the problem of propagating a Gaussian times a first order polynomial.\cite{Lee1982} Zero order terms are treatable within the formalism up to this point, since constants commute with the propagator. Linear contributions are resolved in the following way,
\begin{align}
    U_t c_1^T (q-q_0))\phi_0 &= U_t c_1^T \frac{\hbar}{i}\nabla_{p_0} \phi_0,\\
    &= c_1^T \frac{\hbar}{i}\nabla_{p_0} U_t\phi_0\\
    &= c_1^T \frac{\hbar}{i}\nabla_{p_0} \phi_t\\
    &= c_1^T(M^T_{pp,t} -M^T_{qp,t}A_t)(q-q_t) \phi_t
\end{align}

where the fact was used that the propagator within the LHA commutes with the derivative with respect to the initial momentum $[U_t,\nabla_{p_0}]=0$.\cite{Patoz2018,BEGUSIC2018_3TGA} The derivative with respect to the initial momentum $\nabla_{p_0}$ introduces new terms due to the dependency of the time-dependent parameters of the Gaussian wave packet on the initial conditions of the classical trajectory and necessitates the calculation of the monodromy matrix\cite{Huber1988,Farantos1992,Zhuang2012} \(M_t\), defined by
\begin{equation}
M_t= \left( \begin{matrix}  M_{qq} & M_{qp} \\ M_{pq} & M_{pp} \end{matrix} \right) =   \left( \begin{matrix}  \frac{\partial q_t}{\partial q_0} & \frac{\partial q_t}{\partial p_0 } \\ \frac{\partial p_t}{\partial q_0} & \frac{\partial p_t}{\partial p_0} \end{matrix} \right) \label{eq:monodromy_definition}, \
\frac{d}{dt} M_t =\left( \begin{matrix}  0 & 1 \\ -V''_t & 0 \end{matrix} \right) M_t,
\end{equation}
with \( \left( \frac{\partial q_t}{\partial q_0} \right)_{i,j}  = \frac{\partial q_i(t,q_0,p_0)}{\partial q_{0,j}}\,. \)
In the case of electronic dipole transitions $c_0=\mu_\alpha(q_0)$ and $c_1= \nabla_q \mu_\alpha|_{q_0}$, where $\alpha \in \{x,y,z\}$, corresponding to the  Franck-Condon and Herzberg-Teller terms respectively. The coefficients for nonadiabatic transitions are $c_0=\frac{i}{\hbar}p_0$ and $c_1=-\frac{i}{\hbar}\tau A_0$,  obtained by evaluating $-\tau \nabla_q \phi_0$. The trajectory, Hessians and the time-dependent parameters need to be computed only once as none of the time-dependent parameters depend on $c_0$, $c_1$ or $\alpha$. This enables an efficient treatment of rotational averaging and the combined calculation of emission and internal conversion.
%This approach has been called Extended Thawed Gaussian Approximation (ETGA), since it expands the applicability of TGA.

\subsection{Canonical Ensemble}

The methodology is readily transferred to the thermal problem,\cite{Begusic2020,Begusic2021} all that is needed is a  modification due to the sign difference of the final and initial vibronic Hamiltonian as defined in Eq. \ref{eq:thermal_vibronic_hamiltonian} and a doubling of the $N$ degrees of freedom $q \rightarrow \bar q = (q,q')$ and $p \rightarrow \bar p=(p,p')$, where the primed coordinates are associated with the initial electronic state and the unprimed ones with the final electronic state. To this end, we introduce the $2N \times 2N$ dimensional matrix 
\begin{equation}
W=\left(\begin{matrix}
    1 & 0 \\
    0 & -1 
\end{matrix}\right),\end{equation}
where $1$ stands for a $N\times N$ identity matrix. The vibronic Hamiltonian for the thermal wave packet can then be written as
\begin{equation}
    \bar H_t(\bar q) = -\frac{\hbar^2}{2}\nabla^T_{\bar q}W\nabla_{\bar q} + \bar V_t(\bar q),
\end{equation}
using 
\begin{equation}
    \bar V_t(\bar q) = V_f(q_t) - V_i(q_t'),
\end{equation}
leading to the following set of equations for the time-dependent parameters
\begin{align}
    \dot {\bar q}_t &= W \bar p_t, \quad \dot {\bar p}_t = - \bar V'_t \label{eq:thermal_eom_qp_t}, \\
    \dot {\bar A}_t &= - \bar A_t W \bar A_t - \bar V''_t\label{eq:thermal_eom_A_t}, \\
    \dot {\bar \gamma}_t &= \frac{i\hbar}{2}\Tr( W \bar A_t) + \bar L_t\label{eq:thermal_eom_gamma_t},
\end{align}
with $\bar L_t(\bar q_t, \bar p_t)= L_f(q_t,p_t) - L_i(q'_t, p'_t)$, where $L_{f}(q_t,p_t)$ is the classical Lagrangian associated with a trajectory propagated on the final electronic PES and $L_i(q'_t, p'_t)$ on the initial state PES.  

The monodromy matrix in the thermal case is given by
\begin{equation}
\bar M_t= \left( \begin{matrix}  \bar M_{\bar q \bar q} & \bar M_{\bar q\bar p} \\ M_{\bar p \bar q} & M_{\bar p\bar p} \end{matrix} \right) =   \left( \begin{matrix}  \frac{\partial \bar q_t}{\partial \bar q_0} & \frac{\partial \bar q_t}{\partial \bar p_0 } \\ \frac{\partial \bar p_t}{\partial \bar q_0} & \frac{\partial \bar p_t}{\partial \bar p_0} \end{matrix} \right), \ \frac{d}{dt} \bar M_t =\left( \begin{matrix}  0 & W \\ -\bar V''_t & 0 \end{matrix} \right) \bar M_t. \label{eq:d_thermal_monodromy_matrix_dt}
\end{equation}

The polynomial problem can be solved in the same way as before by replacing all quantities by the thermal counterpart, where the nonadiabatic coupling vector and the transition dipole moment derivative are given by 
\begin{equation}
\tau \rightarrow \bar \tau = \left( \begin{matrix} \tau \\ 0 \end{matrix} \right), \qquad \nabla_q\mu|_{q_0}  \rightarrow \nabla_{\bar q}\mu|_{\bar q_0} = \left( \begin{matrix}
    \nabla_q\mu|_{q_0} \\ 0
\end{matrix} \right).    
\end{equation}

If required, the equations in Cartesian coordinates are  obtained by replacing $q,p$ with the Cartesian position and momentum and a substitution of $\pm 1$ in the matrix $W$ with the inverse mass matrix $\pm M^{-1}$, where $M$ is simply a diagonal matrix of nuclear masses.

\subsection{Initial conditions}

An analytic expression for the density operator of a canonical ensemble is available for harmonic potentials and the initial conditions for equations \ref{eq:thermal_eom_qp_t}--\ref{eq:thermal_eom_gamma_t} can be derived by comparison of the Ansatz at $t_0$ with it.\cite{Schleich2001,Begusic2020} The vibrational density operator of a canonical ensemble in normal coordinates is
\begin{align}
    \rho_\beta(q,q') &= \sqrt{\det  [ \Omega \tanh(\beta \hbar \Omega / 2) /(\pi \hbar )]} \nonumber \\
    &\times \exp\left( \frac{i}{2\hbar}(q,q') A_\beta
    \left( \begin{matrix}
        q \\ q'
    \end{matrix}\right) \right)
\end{align}

with 
\begin{equation}
        A_\beta =  i \left( \begin{matrix}
    \Omega \coth(\beta\hbar \Omega) & -\Omega\sinh(\beta \hbar \Omega)^{-1} \\
    -\Omega\sinh(\beta \hbar \Omega)^{-1} & \Omega \coth(\beta\hbar \Omega)
    \end{matrix}\right).  
\end{equation}
The $N$-dimensional square matrix $\Omega$ is diagonal and contains the angular frequencies of the normal modes of the initial potential surface. The hyperbolic functions of the matrix are understood as diagonal matrices that contain the function evaluated on each diagonal element, and matrix multiplication is implied without explicit dot product notation. $A_\beta$ is a $2N \times 2N$ matrix. 

Based on Eqs. \ref{eq:thermal_phi_0} and \ref{eq:K_acting_on_phi}, the initial conditions have to match  $\rho_\beta^{1/2}$ at $t_0$. It is not necessary to calculate the square root of the density operator explicitly, noting that
\begin{align}
    \rho_\beta^{1/2} &= \rho_{\beta/2} \frac{ Z_{\beta/2} }{Z^{1/2}_\beta}
\end{align}
from which follows that 
\begin{align}
    \bar A_0 &= A_{\beta/2},\\
    \bar q_0 &= \bar 0,   \\
    \bar p_0 &= \bar 0,
\end{align}

where $\bar 0$ is a $2N$-dimensional zero-vector, obtained by comparison of the polynomial term within the exponent of our Ansatz and $\rho^{1/2}_\beta(q,q') \propto \rho_{\beta/2}(q,q')$. 

The missing initial condition for parameter $\bar \gamma_0$ is determined using the fact that the trace of the density operator is normalized. The real part of $\text{Re}(\bar \gamma_0)=0$, as it would only induce a complex phase factor, while the imaginary part of $\bar \gamma_0$ determines the norm of the wave packet and can be derived using the aforementioned trace,
\begin{align}
    \int dq dq' \ \rho_\beta(q,q') = \int dq dq'\  |\rho_\beta^{1/2}(q,q')|^2 = 1.
\end{align}
Using Eq. (\ref{eq:tga_definition}) as Ansatz 
\begin{equation}
\rho_\beta^{1/2}(\bar q) = \exp \left(\frac{i}{\hbar} \left[\frac{1}{2}(\bar q- \bar q_0)^T \bar A_0(\bar q- \bar q_0)+ \bar p^T_0(\bar q-\bar q_0) + \bar \gamma_0 \right] \right)    
\end{equation}
with the normalization condition and $\bar q = (q,q')$ leads to
\begin{align}
    \int dqdq' \ |\phi_{0,\beta}(q,q')|^2 &= 1
\end{align}
which results in 
\begin{align}
    \bar \gamma_0 &= -\frac{i\hbar}{4} \ln\left( \det [\text{Im}\bar A_0/\pi\hbar] \right)\\
    &=-\frac{i\hbar }{2} \ln(\det[\Omega/\pi\hbar]).
\end{align}

The monodromy matrix at $t_0$ is simply a unit matrix since 
\begin{equation}
    \bar M_0 =   \left( \begin{matrix}  \frac{\partial \bar q_0}{\partial \bar q_0} & \frac{\partial \bar q_0}{\partial \bar p_0 } \\ \frac{\partial \bar p_0}{\partial \bar q_0} & \frac{\partial \bar p_0}{\partial \bar p_0} \end{matrix} \right) = \left(\begin{matrix}
    \bar 1 & 0 \\
    0 & \bar 1 
\end{matrix}\right),
\end{equation}
where each $\bar 1$ is a $2N \times 2N$ identity matrix.

\subsection{Improving Computational Efficiency}

Knowledge of the monodromy matrix can be used to solve the equation of motion for the width matrix $\bar A_t$. Showing this requires the definition of two auxiliary time-dependent matrices $\bar Z_t$ and $\bar P_t$, each  with a dimension of $2N \times 2N$. They are connected to the width matrix by the relation
\begin{equation}
    \bar A_t = \bar P_t \bar Z_t^{-1},
\end{equation}

with the following time dependency,\cite{Lee1982,Begusic2019,lasser_lubich_2020}

\begin{equation}
    \frac{d}{dt} \left( \begin{matrix}
        \bar Z_t  \\
        \bar P_t
    \end{matrix} \right) 
    =  \left( \begin{matrix} 
        0 & W \\
        -\bar V''_t & 0
    \end{matrix} \right) \left( \begin{matrix}
        \bar Z_t  \\
        \bar P_t
    \end{matrix} \right),
\end{equation}
which is solved using 
\begin{equation}
    \left( \begin{matrix}
        \bar Z_t  \\
        \bar P_t
    \end{matrix} \right)  = 
    \bar M_t\left( \begin{matrix}
        \bar Z_0  \\
        \bar P_0
    \end{matrix} \right),
\end{equation}

and combined with the initial condition $\bar A_0 = \bar P_0 \bar Z^{-1}_0$. The initial values of this pair of auxiliary matrices are underdetermined, and we are free to set $\bar Z_0=1$ and $\bar P_0 = \bar A_0$. It is noteworthy that the temperature dependence is entirely determined by $\bar A_0$, which means that once the monodromy matrix has been integrated, calculations at different temperatures are readily available requiring only matrix products instead of  the integration of the differential equation of $\bar A_t$ at different temperatures.

% Knowledge of the monodromy matrix can be used to solve the integration of the width matrix $\bar A_t$. Showing this requires the definition of two auxiliary time-dependent matrices $\bar Z_t$ and $\bar P_t$, each square with a dimension of $2N \times 2N$ and with the following time dependency,\cite{Lee1982,Begusic2019,lasser_lubich_2020}

% \begin{equation}
%     \frac{d}{dt} \left( \begin{matrix}
%         \bar Z_t  \\
%         \bar P_t
%     \end{matrix} \right) 
%     =  \left( \begin{matrix} 
%         0 & W \\
%         -\bar V''_t & 0
%     \end{matrix} \right) \left( \begin{matrix}
%         \bar Z_t  \\
%         \bar P_t
%     \end{matrix} \right) 
% \end{equation}
% which is solved using 
% \begin{equation}
%     \left( \begin{matrix}
%         \bar Z_t  \\
%         \bar P_t
%     \end{matrix} \right)  = 
%     \bar M_t\left( \begin{matrix}
%         \bar Z_0  \\
%         \bar P_0
%     \end{matrix} \right) .
% \end{equation}

% The connection to matrix $\bar A_t$ is made by realizing, that
% \begin{equation}
%     \bar A_t = \bar P_t \bar Z_t^{-1}
% \end{equation}
% solves the differential equation for $\bar A_t$ as long as $\bar A_0 = \bar P_0 \bar Z^{-1}_0$. The initial values of this pair of auxiliary matrices are undetermined, and we are free to set $\bar Z_0=1$ and $\bar P_0 = \bar A_0$. It is noteworthy that the temperature dependency is entirely determined by $\bar A_0$, which means that once the monodromy matrix has been integrated, calculations at different temperatures are readily available. 

The auxiliary matrices also enable an analytic integration of a term in Eq. \ref{eq:thermal_eom_gamma_t}, required for $\bar \gamma_t$,  that involves the trace over $\bar A_t$,\cite{lasser_lubich_2020,Begusic2022}
\begin{align}
    \int_0^t dt' \Tr (W \bar A_{t'}) &= \int dt' \Tr (\bar W P_{t'} Z_{t'}^{-1})\\
    &= \int^t_0 d{t'} \Tr \left( \left(\frac{d}{d{t'}}\bar  Z_{t'} \right) \bar Z_{t'}^{-1} \right)\\
    &= \ln \left( \det (\bar Z_t \bar Z_0^{-1}) \right)
\end{align}

It should be mentioned that it is possible to avoid the definition of matrix $\bar A_t$ entirely, using $\bar P_t$ and $\bar Z_t$ from the start in the definition of the Gaussian wave packet, which is known as Hagedorn parametrization that constitutes an equivalent alternative in the formulation of the TGA.\cite{Hagedorn1980,Hagedorn1985,Hagedorn1998}

The thermal problem seems to require the numeric integration of equation \ref{eq:d_thermal_monodromy_matrix_dt}, a $4N\times 4N$ dimensional problem. Using the harmonic approximation for the initial potential and rearrangement of the coordinate indexing enables the splitting of the problem into a $2N \times 2N$ integral, and a set of $N$ $2\times2$ dimensional integrals that can be solved analytically. 

Expanding equation  \ref{eq:d_thermal_monodromy_matrix_dt} in the original coordinates yields,

\begin{align}
   \frac{d}{dt} \bar M_t &= \left( \begin{matrix}  0 & 0 & 1 & 0 \\ 0 & 0 & 0  & -1 \\ -V''_f(q_t) &0 &0 &0 \\ 0 & V''_i(q'_t) &0 &0  \end{matrix} \right) \bar M_t.
\end{align}
This form corresponds to an order of coordinates given by $(\bar q, \bar p) = (q,q',p,p')$. The separable block form becomes apparent when the order of coordinates is changed to $(q,p,q',p')$, effectively swapping the third and fourth row and the second and third column, such that
\begin{align}
   \frac{d}{dt} \bar M_t &= \left( \begin{matrix}  0 & 1 & 0 & 0 \\ -V''_f(q_t) & 0 & 0  & 0 \\ 0 &0 &0 & -1 \\ 0 & 0 & V''_i(q'_t) &0 \end{matrix} \right) \bar M_t
\end{align}
There is no mixing of the coordinates associated with the initial and the final potential, and the set of differential equations associated with either potential can be solved independently,

\begin{equation}
O_{f} = \left( \begin{matrix}  0 & 1 & \\ -V''_{f}(q_t) & 0 \end{matrix} \right), O_{i} = \left( \begin{matrix}  0 & -1 & \\ V''_{i}(q_t') & 0 \end{matrix} \right)\\  
\end{equation}

\begin{equation}
\frac{d}{dt}\bar  M_t = \dot M_{t,f} \oplus \dot M_{t,i} = O_f M_{t,f} \oplus O_{t,i} M_{t_i}.
\end{equation}

Further simplifications are possible for the initial potential part. The initial state trajectory remains at rest, since we start at the equilibrium position with zero momentum as initial condition. This simplifies the LHA of the initial potential to an ordinary time-independent harmonic potential obtained by second order expansion around the initial state equilibrium position. The second derivative is then simply a square $N\times N$ matrix with the squared normal mode frequencies on its diagonal, $ V''_i(q'_t) = V''_i(q'_{eq}) = \Omega^2_i$. Rearranging the coordinates $(q',p')$ to $(q'_1,p'_1, \dots, q'_N, p'_N )$ reduces the matrix equation for the initial state monodromy matrix to a direct sum of $N$ analytically solvable independent ordinary linear differential equations of dimension $2\times 2$, 
\begin{equation}
    \dot M_{t,i} = O_{t,i}M_{t,i} = \oplus_n^N \left( \begin{matrix} 0 & -1 \\ \omega^2_{ni}  & 0 \end{matrix}  \right) M_{t,ni}.
\end{equation}
This shows that the initial state part of the thermal problem is trivial and requires no effort beyond a common geometry optimisation and frequency calculation of the initial electronic state. 

The expensive part is the propagation of a trajectory on the final state potential, starting at the equilibrium geometry of the initial state and the calculation of the Hessians along this trajectory to obtain $V''_f(q_t)$ for the integration of the final state monodromy matrix $M_{t,f}$.

% Passt eher in eine Introduction als in den Theorie Teil.
% Up to this point we have made no approximations regarding the potential energy surfaces. But treating more than a few dimensions using exact potentials is  prohibitively expensive. There are a few options available to tackle this problem. One way is limiting the degrees of freedom to those which are assumed to be responsible for the transitions. Another way is using approximated potentials, that enable efficient solutions to the pertinent equations, including all degrees of freedom. In this work we take the latter approach, using a semi-classical approach in tandem with the harmonic approximation. 

% Applying it to the initial and final potential enables an analytic solution for canonical ensemble time-dependent correlation function, which includes the temperature free case as limit as $T$ goes to zero. The adiabatic harmonic (AH) and vertical harmonic (VH) model are based on the harmonic expansion of the true potential either around its optimized geometry or at the geometry of the optimized initial state in the vertical case. 

\section{Computational Details}

Reliable and robust methods are desirable for theoretical and computational chemistry, but accurate predictions without adjustments based on external experimental parameters remain challenging.
The search for a robust "black-box" approach motivated the use of a straightforward protocol, applied to all calculations, and the ambition to treat molecular systems with all nuclear degrees of freedom included, led to the choice of density functional theory as electronic structure method. 

The Gaussian16\cite{g16} program package was used to optimize structures and to obtain gradients and Hessians, at the DFT and TD-DFT level. The  B3LYP\cite{Becke1993,Stephens1994}, CAM-B3LYP\cite{Yanai2004}, PPBE0\cite{Adamo1999,Ernzerhof1999} and $\omega$B97XD\cite{Chai2008} functionals, together with a 6-311G basis set, have been applied to formaldehyde, fluorobenzene and azulene. In the case of the largest molecule, the squaraine dye, calculations were limited to the B3LYP and CAM-B3LYP functionals. 
The software package also provided transition dipole moments, transition dipole momenent gradients and the nonadiabatic coupling vector, between the excited and the ground state. In the case of azulene, the nonadiabatic coupling had to be calculated between the second and first excited state, which was obtained with the help of the Q-Chem\cite{Shao2015} package, using the auxiliary wave function approach.\cite{Zhang2014,Ou2014,Wang2021a}

Trajectories required for the ETGA dynamics were propagated for 150 fs using a 0.05 fs step, Hessians were calculated at each step. The Cartesian coordinates of the trajectory and the Hessians were transformed to the normal coordinate system of the initial electronic state for the integration of the differential equations and the calculation of the auto correlation functions.
The FCclasses3\cite{Cerezo2023} program was used to execute all vertical and adiabatic harmonic model calculations,\cite{Hazra2003,Hazra2004,Ferrer2012} using the time-dependent approach, to obtain the correlation function for a total time of 200.0 fs, using 8192 steps. 
Gaussian envelope functions with half width at half maximum (HWHM) values of 0.025, 0.05 and 0.1 eV have been used in all calculations, to avoid Gibbs artifacts due to the Fourier transform of the finite time domain signal and to test the dependency of the results on the choice of the lineshape function (LSF). A Lorentzian with a HWHM of 0.001 eV had to be used in conjunction with the Gaussian LSF's, yielding a Voigt profile, for fluorobenzene and the squaraine dye to obtain reasonable internal conversion rates.

\section{Formaldehyde}

\begin{table}
\centering
\caption{\label{tab:formaldehyde_results} Energy gap, spontaneous emission rate, internal conversion rate and quantum yield using different DFT functionals and models for formaldehyde $S_1 \rightarrow S_0$. The values are averaged over calculations using Gaussian lineshape functions with hwhm values of [0.025, 0.05, 0.1] eV. The mean values are tabulated, with the standard deviation in parentheses. Temperature for ETGA calculation 293.15K. The experimental values are taken from ref.~\onlinecite{Miller1978}, gasphase at room temperature.}
\begin{ruledtabular}
\begin{tabular}{lrccc}  
$\Delta E / \mathrm{eV}$ & EXP = 3.49 & AH\footnote{Adiabatic energy difference of optimized electronic states} & ETGA\footnote{Vertical energy difference at the optimized geometry of the initial state} & VH\footnote{Adiabatic energy difference between the initial state and harmonic approximation for the final state, based on the gradient and hessian at the optimized geometry of the initial state.} \\
& B3LYP     & 3.46 & 2.96 & 3.60\\
& CAM-B3LYP & 3.48 & 3.16 & 3.56\\
& PBE       & 3.49 & 3.04 & 3.61\\
& $\omega$B97XD    & 3.51 & 3.11 & 3.61\\
\hline
$k_{\textrm{SE}} /$     & EXP = 3.049 & AH         & ETGA       & VH \Tstrut \\
$[10^5 \textrm{s}^{-1}]$  & B3LYP     & 1.28(0.00) & 1.97(0.00)  & 1.51(0.00)\\
                          & CAM-B3LYP & 1.41(0.00) & 1.99(0.00) & 1.74(0.00)\\
                          & PBE       & 1.34(0.00) & 2.10(0.00)  & 1.57(0.00)\\
                          & $\omega$B97XD    & 1.32(0.00) & 2.07(0.00)  & 1.54(0.00)\\
\hline
$k_{\textrm{IC}} /$ &  EXP = 1.189 & AH & ETGA & VH\Tstrut \\
$[10^7 \textrm{s}^{-1}]$ &B3LYP     & 3.598(0.033) & 0.240(0.005) & 0.266(0.004)\\
                         &CAM-B3LYP & 9.748(0.061) & 7.354(0.101) & 6.760(0.726)\\
                         &PBE       & 3.822(0.029) & 0.455(0.007) & 0.482(0.008)\\
                         &$\omega$B97XD    & 3.403(0.019) & 0.803(0.007) & 0.829(0.012)\\
\hline
$\Phi_\text{QY}$ & EXP = 2.5        & AH           & ETGA         & VH \Tstrut \\
$\times 10^2$            &B3LYP     & 0.356(0.003) & 7.610(0.157) & 5.377(0.081)\\
                         &CAM-B3LYP & 0.144(0.001) & 0.269(0.004) & 0.260(0.030)\\
                         &PBE       & 0.350(0.003) & 4.413(0.064) & 3.143(0.048)\\
                         &$\omega$B97XD    & 0.387(0.002) & 2.516(0.022) & 1.825(0.025)\\
\hline
\end{tabular}
\end{ruledtabular}
\end{table}

 \begin{figure}
  \centering
  \includegraphics[width=.50\textwidth, trim=0 0.25cm 0 0.5cm, clip]{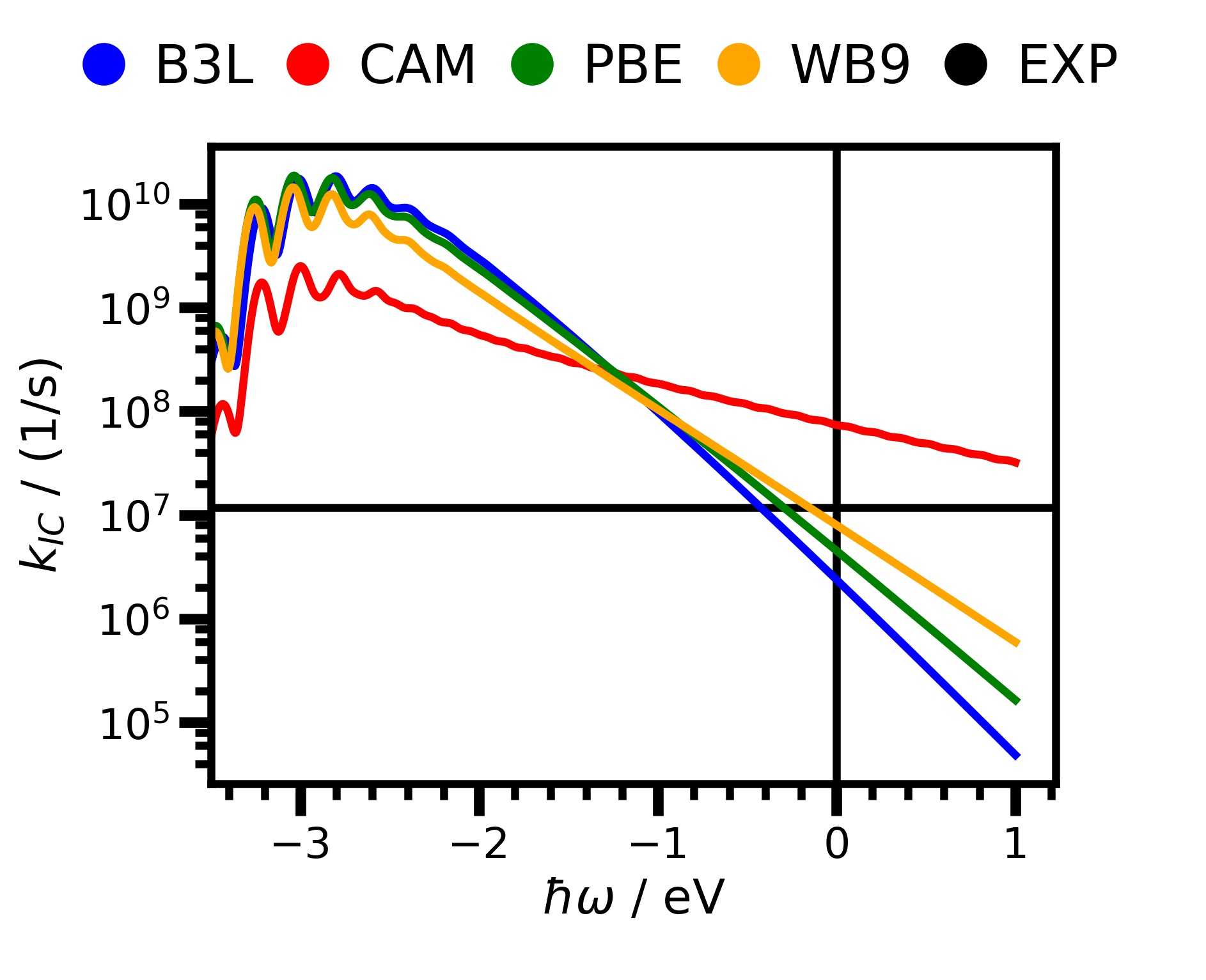}
\vspace{-20pt}
  \caption{ETGA internal conversion spectrum of the $S_1$ state of \textbf{formaldheyde}, calculated using DFT/TDDFT with the B3LYP, CAM-B3LYP, PBE and $\omega$B97XD functionals. The horizontal black line indicates the experimental rate, the vertical black line the point of energy conservation. The spectrum is based on a thermal correlation function (T=298.15 K) of 150.0 fs, broadened with a Gaussian line shape function with a HWHM value of 0.05 eV.}
  \label{fig:etga_kic_formaldheyde_functionals}
\end{figure}

 Radiative and nonradiative lifetime values of the first excited singlet state of formaldehyde have been reported, resolved for vibronic states and followed by measurements of rates dependent on the initially populated rovibronic state.\cite{Jeunehomme1964,Sakurai1971,Aoki1973,Yeung1973,Miller1975,Miller1976,Shibuya1978a,Molina1978,Shibuya2008,Weisshaar1979} The dependency of the internal conversion rate on the initial vibrational state was also investigated in a computational study, using a  Herman-Kluk (HK) frozen Gaussian Semiclassical Initial Value Representation propagator.\cite{Ianconescu2011,Ianconescu2013,Herman_1984,Kluk1986} The experimental excitation energy of the vibrational ground state of the first excited state is reported at 3.49 eV and it is taken as reference value for the adiabatic energy difference of the optimized S$_1$ and S$_0$ state.\cite{Miller1978} This energy difference is well reproduced by all functionals, and the particular values for each functional are given in Table \ref{tab:formaldehyde_results} (top section $\Delta E$, column AH). The vertical energy difference of the electronic states is given in column ETGA. The last column VH lists also an adiabatic energy difference, but in this case the energy of the ground state is derived from a harmonic approximation of the ground state potential using the optimized $S_1$ geometry as expansion point. Differences of the AH and VH value indicate anharmonicity of the real potential, since a perfectly harmonic potential would yield the same value regardless of the expansion point. 

 All spontaneous emission rate calculations, listed in Table \ref{tab:formaldehyde_results}, underestimate the experimental value, but the emission  rates are quite consistent within each model for the various functionals and there is a clear tendency between the dynamical models. The adiabatic harmonic model yields the lowest rates, followed by the vertical model which is exceeded by the ETGA results, which are roughly two-thirds of the reported value. The internal conversion rates show a different picture. The adiabatic model consistently overestimates the experimental rate. ETGA and VH model yield results that are below the experimental value, except for CAM-B3LYP which overestimates the rate. The internal conversion rate of the ETGA model increases by a factor of $\approx$31 when the electronic structure back bone is changed from B3LYP to CAM-B3LYP. The other functionals do not induce such a drastic change and the values remain within a factor of four. The VH model values are similarly affected. The variation of the rates with the half-width-at-half-maximum value of the Gaussian lineshape function is small and the results are stable in this regard. The internal conversion rate spectrum using the ETGA is shown in Fig. \ref{fig:etga_kic_formaldheyde_functionals}, to ensure that the results for the internal conversion rate are not due to some accidental agreement.

 The experimental fluorescence quantum yield of 2.5 percent is accurately predicted by the ETGA method in combination with the $\omega$B97XD functional. But this is clearly a case of fortuitous error cancellation as the emission rate and the internal conversion rate are both approximately wrong by a factor of two-thirds, which cancel each other out when the quantum yield is calculated. The AH model on the other hand "suffers" twice, as it underestimates the emission rate while overestimating the internal conversion rate, leading to worst results for fluorescence quantum yields of the $S_1$ state of formaldehyde. The VH results are close the ETGA model and close to the experimental value, but again due to an error cancellation. 

 Disentangling the error into contributions due to the dynamical and the electronic structure methods is difficult, as the internal conversion spectrum cannot be observed directly. But for emission experiment and theory can be compared. All emission spectra using the smallest HWHM of 0.025 eV for the broadening are given in the supplementary material, Fig. S1. It is clear that the detailed vibronic structure is not accurately described by any of the models, but the ETGA model is an improvement as it captures the outline better and avoids a lengthy vibrational progression into the low energy region. Among the functionals CAM-B3LYP stands out as worst option, there even the ETGA model predicts an erroneous progression alike to the AH and VH model. In the case of formaldehyde it appears likely that both aspects contribute to the deviation from the experiment, the static electronic structure calculations and the dynamical models.

% It is difficult to answer why the adiabatic model yields such a high internal conversion rate. It is not due to the nonadiabatic coupling vector or the projection of the vector on normal modes. Indeed, the opposite effect would be expected since the value of the initial projection is the lowest for CAM-B3LYP. But a plot of the internal conversion rate against the energy gap indicates a slower decrease of the rate with energy, which leads to a higher rate at the point of degeneracy of the initially populated states and the approximately isoenergetic vibrational states of the ground state potential.

\section{Fluorobenzene}

\begin{table}
\centering 
\caption{\label{tab:fluorobenzene_results} Energy gap, spontaneous emission rate, internal conversion rate and quantum yield using different DFT functionals and models for fluorobenzene $S_1 \rightarrow S_0$. The values are averaged over calculations using Gaussian lineshape functions with hwhm values of [0.025, 0.05, 0.1] eV combined with a Lorentzian with a fixed hwhm value of 0.001 eV. The internal conversion rate varies tremendously with the width of the Lorentzian lineshape function, and is $\approx 0$ if only Gaussian broadening is used. The mean values are tabulated, with the standard deviation in parentheses.}
\begin{ruledtabular}
\begin{tabular}{lrccc}  
$\Delta E / \mathrm{eV}$ & EXP = 4.69 & AH\footnote{Adiabatic energy difference of optimized electronic states} & ETGA\footnote{Vertical energy difference at the optimized geometry of the initial state} & VH\footnote{Adiabatic energy difference between the initial state and harmonic approximation for the final state, based on the gradient and hessian at the optimized geometry of the initial state.} \\
& B3LYP     & 5.33 & 5.18 & 5.34\\
& CAM-B3LYP & 5.50 & 5.34 & 5.50\\
& PBE       & 5.46 & 5.31 & 5.47\\
& $\omega$B97XD    & 5.48 & 5.32 & 5.49\\
\hline
$k_{\textrm{SE}} /$ & EXP = 3.51 & AH & ETGA & VH \Tstrut \\
$[10^7 \textrm{s}^{-1}]$  & B3LYP     & 1.19(0.00) & 1.34(0.00) & 1.44(0.00)\\
                          & CAM-B3LYP & 1.42(0.00) & 1.53(0.00) & 1.65(0.00)\\
                          & PBE       & 1.42(0.00) & 1.59(0.00) & 1.72(0.00)\\
                          & $\omega$B97XD    & 1.37(0.00) & 1.52(0.00) & 1.64(0.00)\\
\hline
$k_{\textrm{IC}} /$ &  EXP = 6.79 & AH & ETGA & VH\Tstrut \\
$[10^7 \textrm{s}^{-1}]$ &B3LYP     &  1.95(0.00) & 2.18(0.00) &  2.12(0.00)\\
                         &CAM-B3LYP &  2.16(0.00) & 2.42(0.00) &  2.35(0.00)\\
                         &PBE       &  1.94(0.00) & 2.17(0.00) &  2.10(0.00)\\
                         &$\omega$B97XD    &  2.12(0.00) & 2.37(0.00) &  2.31(0.00)\\
\hline
$\Phi_\text{QY}$ & EXP = 34.10 & AH & ETGA & VH \Tstrut \\
$\times 10^2$ &B3LYP     & 37.91(0.01) &  38.04(0.01) & 40.49(0.01)\\
              &CAM-B3LYP & 39.58(0.01) &  38.77(0.01) & 41.30(0.01)\\
              &PBE       & 42.37(0.01) &  42.36(0.01) & 44.99(0.01)\\
              &$\omega$B97XD    & 39.30(0.01) &  38.98(0.01) & 41.51(0.01)\\
\hline
\end{tabular}
\end{ruledtabular}
\end{table}

 \begin{figure}
  \centering
  \includegraphics[width=.50\textwidth, trim=0 0.25cm 0 0.5cm, clip]{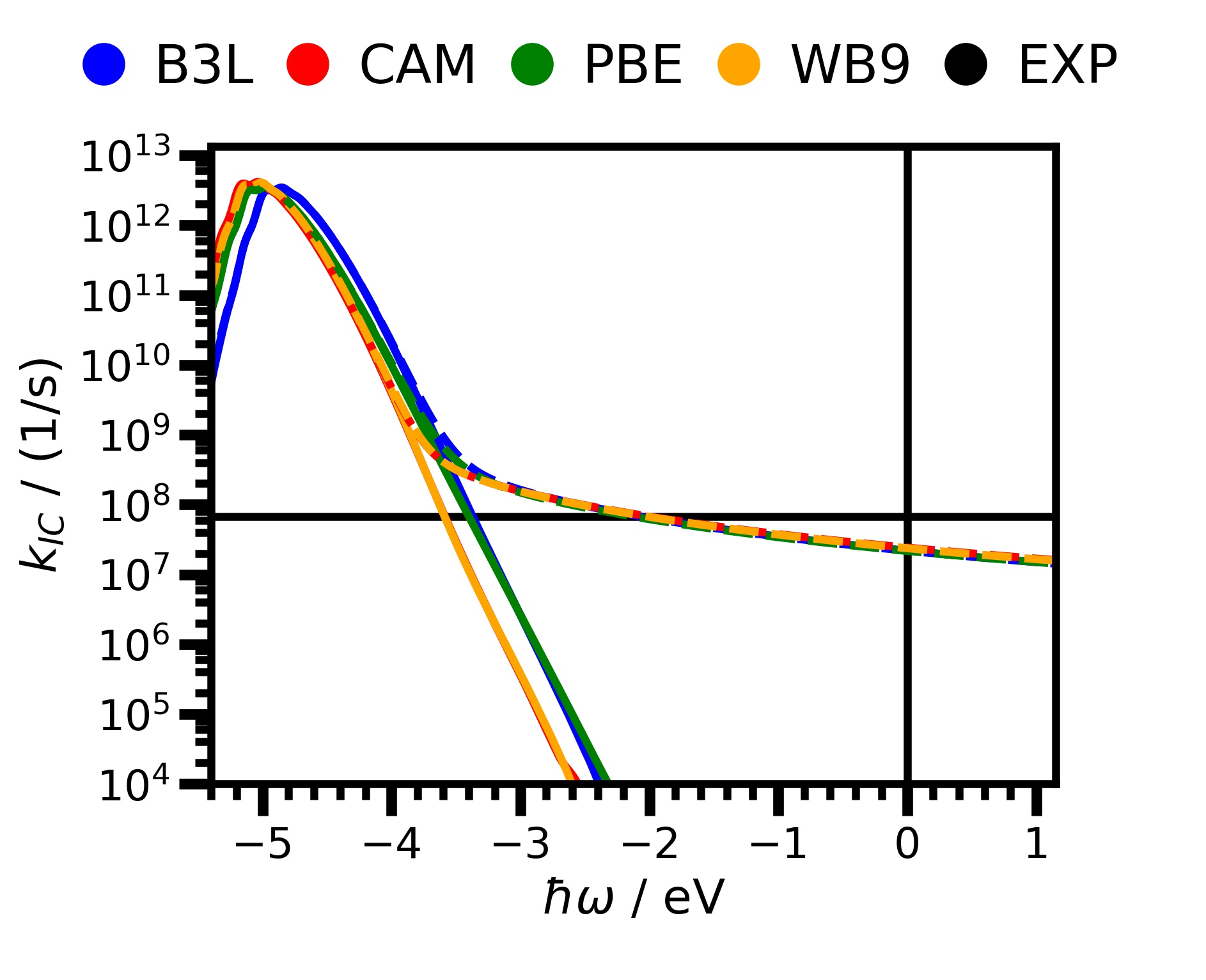}
\vspace{-20pt}
  \caption{ETGA internal conversion spectrum of the $S_1$ state of \textbf{fluorobenzene}, calculated using DFT/TDDFT with the B3LYP, CAM-B3LYP, PBE and $\omega$B97XD functionals. The horizontal black line indicates the experimental rate, the vertical black line the point of energy conservation. The spectrum is based on a thermal correlation function (T=298.15 K) of 150.0 fs. The solid lines show the spectrum obtained using only a Gaussian line shape function with a HWHM value of 0.05 eV, the dashed lines show the spectrum using additionally a Lorentzian broadening function with a HWHM value of 0.001 eV.}
  \label{fig:etga_kic_fluorobenzene_functionals}
\end{figure}

Fluorobenzene is interesting as test system due to its high harmonicity combined with a large energy gap.\cite{Koese2022} The computational results are gathered in Table \ref{tab:fluorobenzene_results} and compared with experimental gas-phase measurements\cite{Abramson1972} at room temperature. The choice of DFT and TD-DFT as backbone for the electronic structure calculations was based on their good and consistent performance for geometries and frequencies of the ground and excited state.\cite{Koese2022} But the performance for energies is worse and the adiabatic energy gap is overestimated considerably by all functionals with deviations ranging from 0.64 to 0.81 eV. The fluorescence rates on the other hand are underestimated, yielding only 34 \% to 49 \% of the experimental value. The VH model predicts the highest values, followed by the ETGA and lastly the AH method. The ETGA and VH model switch places when it comes to  internal conversion rates although the differences between both models are marginal. All models also underestimate the internal conversion rate and it is again a similar factor as in the spontaneous emission rates, which leads to computational quantum yields that appear to be in good agreement with the experimental value, an agreement that is again due to a cancellation of a common deviation in the underestimates of the radiative and nonradiative rates. A theoretical study\cite{He2011} of the internal conversion rate of fluorobenzene, using the displaced harmonic oscillator model and the B3LYP functional, tested dephasing widths for the Lorentzian line shape function ranging from 5  to 10 \CM . The best agreement with the experiment was found to be around 10 \CM, which is almost the same as the 0.001 eV $\approx$ 8 \CM \ used in this work. For fluorobenzene the additional Lorentzian broadening is essential for the prediction of the internal conversion rate, as a pure Gaussian broadening leads to an internal conversion rate of zero. The tremendous effect of the broadening is shown in Fig. \ref{fig:etga_kic_fluorobenzene_functionals}, where the internal conversion spectrum is shown using only a Gaussian broadening of 0.05 eV (solid lines in Fig. \ref{fig:etga_kic_fluorobenzene_functionals}) and the results including a Lorentzian broadening (dashed lines in Fig. \ref{fig:etga_kic_fluorobenzene_functionals}) of 0.001 eV. In this case, the line shape function is clearly the primary factor for the prediction of the internal conversion rate in this case, while the choice of the model appears to be secondary with only a minor effect on the rate.

Comparison of experimental and calculated emission spectra, given in the supplementary material Fig. S2, shows that the vibronic progression is reasonably described by all three models, even though the error in the energy gap is quite large. The propagation time limits the resolution, so it is not clear whether the finer features could be resolved or not, but in this case a main source of error is clearly the electronic structure method, which drastically overestimates the energy gap while the coupling elements appear to be underestimated in magnitude.

\section{Azulene}

\begin{table}
\centering
\caption{\label{tab:azulene_results} Energy gap, spontaneous emission rate, internal conversion rate and quantum yield using different DFT functionals and models for azulene $S_2 \rightarrow S_0$ emission. The spontaneous emission is calculated for the $S_2\rightarrow S_0$ transition, while the internal conversion rate is based on the $S_2 \rightarrow S_1$ transition, as the nonradiative rate $S_2\rightarrow S_0$ is negligible. The values are averaged over calculations using Gaussian lineshape functions with HWHM values of [0.025, 0.05, 0.1] eV. The mean values are tabulated, with the standard deviation in parentheses.}
\begin{ruledtabular}
\begin{tabular}{lrccc}  
$\Delta E / \mathrm{eV}$ & EXP = 3.57; 1.79\footnote{Adiabatic S2-S0 and S2-S1 energy gap} & AH\footnote{Adiabatic energy difference of optimized electronic states} & ETGA\footnote{Vertical energy difference at the optimized geometry of the initial state} & VH\footnote{Adiabatic energy difference between the initial state and harmonic approximation for the final state, based on the gradient and hessian at the optimized geometry of the initial state.} \\
 & B3LYP     & 3.65;1.59 & 3.57;1.42 & 3.65;1.61\\
 & CAM-B3LYP & 3.85;1.72 & 3.79;1.55 & 3.85;1.74\\
 & PBE       & 3.76;1.65 & 3.68;1.47 & 3.76;1.66\\
 & $\omega$B97XD    & 3.84;1.71 & 3.77;1.55 & 3.84;1.73\\
\hline
$k_{\textrm{SE}} /$ & EXP = 1.31 & AH & ETGA & VH \Tstrut \\
$[10^7 \textrm{s}^{-1}]$ &B3LYP     & 1.16(0.00)  & 1.04(0.00) & 1.09(0.00) \\
                         &CAM-B3LYP & 2.02(0.00)  & 2.20(0.00) & 2.28(0.00) \\
                         &PBE       & 1.40(0.00)  & 1.24(0.00) & 1.30(0.00) \\
                         &$\omega$B97XD    & 2.12(0.00)  & 2.17(0.00) & 2.24(0.00) \\
\hline
$k_{\textrm{IC}} /$ &  EXP = 2.99 & AH & ETGA & VH\Tstrut \\
$[10^8 \textrm{s}^{-1}]$ &B3LYP     &  22.47(2.41) & 1.83(0.19) & 1.84(0.34) \\
                         &CAM-B3LYP &  25.93(2.08) & 5.81(0.55) & 6.38(0.62)\\
                         &PBE       &  15.39(1.60) & 1.61(0.16) & 1.56(0.27) \\
                         &$\omega$B97XD    &  23.55(1.94) & 5.22(0.52) & 5.74(0.58) \\
\hline
$\Phi_\text{QY}$ & EXP = 4.20 & AH & ETGA & VH \Tstrut \\
$\times 10^2$ &B3LYP     & 0.52(0.05) & 5.42(0.50)  & 5.74(0.91) \\
              &CAM-B3LYP & 0.78(0.06) & 3.68(0.32)  & 3.49(0.31)\\
              &PBE       & 0.91(0.09) & 7.20(0.62)  & 7.86(1.14)\\
              &$\omega$B97XD    & 0.90(0.07) & 4.02(0.37)  & 3.79(0.35)\\
\hline
\end{tabular}
\end{ruledtabular}
\end{table}

 \begin{figure}
  \centering
  \includegraphics[width=.50\textwidth, trim=0 0.25cm 0 0.5cm, clip]{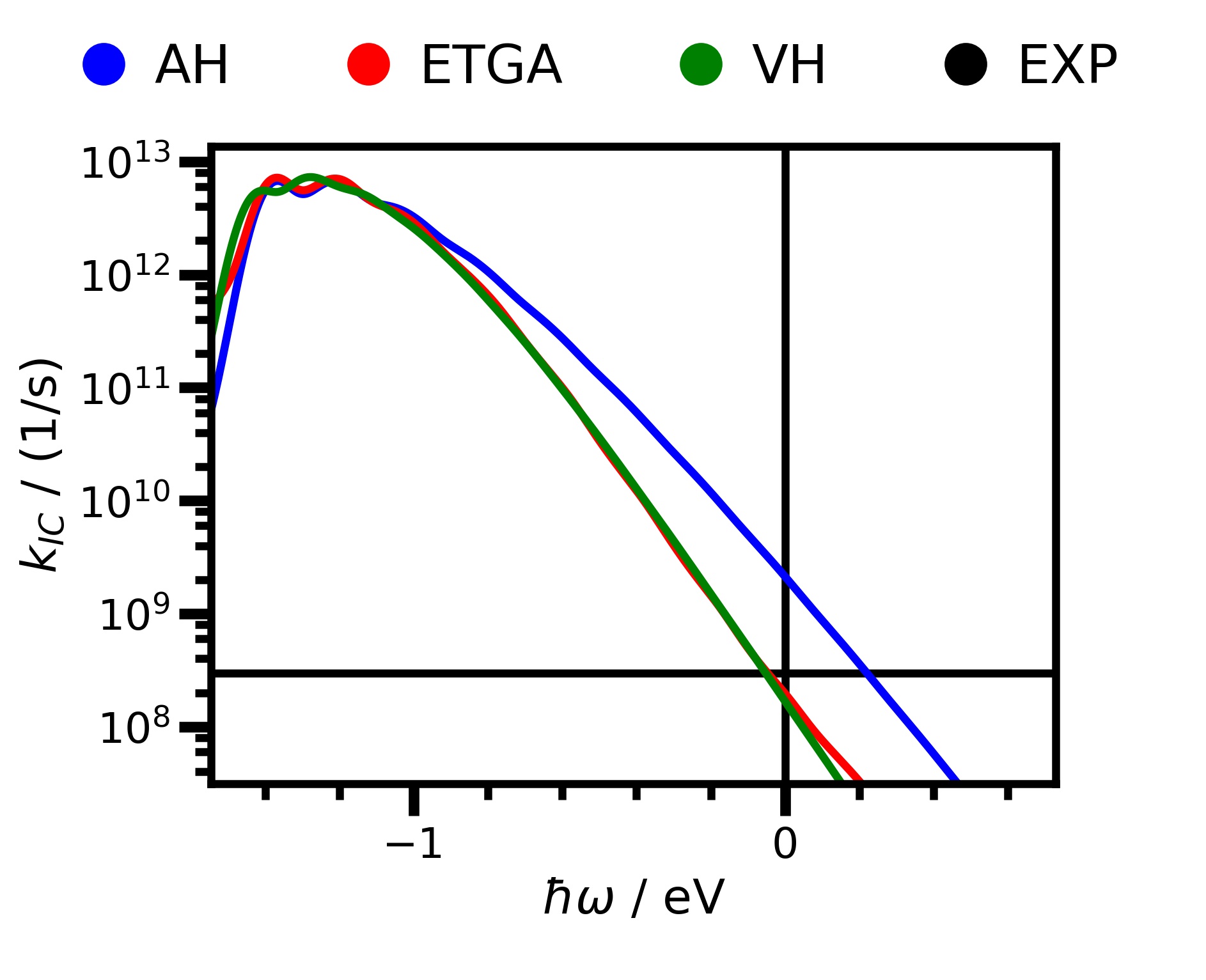}
\vspace{-20pt}
  \caption{Internal conversion spectrum of the $S_2\rightarrow S_1$ transition of \textbf{azulene}, calculated with the adiabatic harmonic model (AH), the extended thawed Gaussian approach (ETGA) and the vertical harmonic model (VH), using the B3LYP functional. The horizontal black line indicates the experimental rate, the vertical black line the point of energy conservation. The spectrum is based on a thermal correlation function (T=298.15 K) of 150.0 fs, using a Gaussian line shape function with a HWHM value of 0.05 eV.}
  \label{fig:azule_b3lyp_kic_spectrum}
\end{figure}

The third system under investigation is azulene, a well-known exception to Kasha's rule,\cite{Kasha1950} with strong $S_2$ to $S_0$ fluorescence, compared to weak emission originating from the $S_1$ state. The experimental values of the quantum yield, radiative and nonradiative rates have been investigated by multiple groups in the past and it has been established that $S_2\rightarrow S_1$ internal conversion is the major competing process, while intersystem crossing can be neglected.\cite{Birks1972,Murata1972,Eber1977,Gillispie1979,Griesser1980} In this work we use gas phase experiments as reference,\cite{Demmer1987,Woudenberg1988} which are comparable to reported values of azulene in various solvents,\cite{Wagner1992} the theoretical results are gathered in Table \ref{tab:azulene_results}. The adiabatic energy gaps are in good agreement with the experimental values for all functionals. The emission rates show little variation with regard to the model, but there is a difference of a factor of approximately two with respect to the choice of the DFT functional. This ratio is reflected in the magnitudes of the predicted transition dipole moments, which are $|\vec \mu^\mathrm{B3LYP}_{20}|^2=0.046$ au, $|\vec \mu^\mathrm{CAM-B3LYP}_{20}|^2=0.091$ au, $|\vec \mu^\mathrm{PBE}_{20}|^2=0.047 $ au and $|\vec \mu^\mathrm{\omega B97XD}_{20}|^2=0.100$ au. The internal conversion rate  however shows a clear distinction of the VH and ETGA model, which are in agreement, and the adiabatic harmonic model, which strongly overestimates the rate, as illustrated in Fig. \ref{fig:azule_b3lyp_kic_spectrum}, depicting the internal conversion spectrum calculated with the three different models combined with the B3LYP functional.

The quantum yields obtained with the ETGA and VH model fluctuate between 3.8 \% and 7.9 \%, close to the experimental value of 4.2 \%, while the results of the adiabatic harmonic model fall within 0.5 to 0.9 \%. The closest result for the quantum yield is 4.0 \% and is obtained with ETGA coupled with the $\omega$B97XD functional, but the agreement stems again from an error cancellation, as this choice overestimates both rates by a similar factor.
The emission spectra of the $S_2\rightarrow S_0$ transition are given in Figure S3 of the supplementary material. The results resemble the reported emission spectrum from reference \onlinecite{Prlj2020}, where it was shown that a single-Hessian\cite{Begusic2019} ETGA is sufficient for the calculation of emission and absorption spectra of azulene. The different dynamical models show some differences in the relative peak intensities when compared to the experiment, in particular the VH model, which overestimates the 0-0 transition but errors in energy and in magnitude of transition dipole elements appear to have a greater effect on the results in this case.

% placing the quantum yield of fluorescence within $\approx$0.040--0.052, the rate of fluorescence within $1.31\textrm{--}4.5 \times 10^7 \ \mathrm{s}^{-1}$ and the nonradiative rate within $2.99$--$9.16 \times 10^8 \ \mathrm{s}^{-1}$.

\section{Squaraine Dye}

\begin{table}
\centering
\caption{\label{tab:sqme_results} Energy gap, spontaneous emission rate, internal conversion rate and quantum yield using different DFT functionals and models for squaraine dye I $S_1 \rightarrow S_0$. The values are averaged over calculations using Gaussian lineshape functions with hwhm values of [0.025, 0.05, 0.1] eV combined with a Lorentzian with a fixed hwhm value of 0.001 eV. The values of the IC rate without Lorentzian broadening are on the order of $\approx 10^4$, which is about four orders of magnitude too small. The mean values are tabulated, with the standard deviation in parentheses.}
\begin{ruledtabular}
\begin{tabular}{lrccc}  
$\Delta E / \mathrm{eV}$ & EXP = 1.776 & AH\footnote{Adiabatic energy difference of optimized electronic states} & ETGA\footnote{Vertical energy difference at the optimized geometry of the initial state} & VH\footnote{Adiabatic energy difference between the initial state and harmonic approximation for the final state, based on the gradient and hessian at the optimized geometry of the initial state.} \\
& B3LYP     & 1.825 & 1.771 & 1.826\\
& CAM-B3LYP & 1.904 & 1.855 & 1.905\\
\hline
$k_{\textrm{SE}} /$ & EXP = 1.875 & AH & ETGA & VH \Tstrut \\
$[10^8 \textrm{s}^{-1}]$  & B3LYP     &  1.295(0.00) & 1.290(0.00) & 1.306(0.00)\\
                          & CAM-B3LYP &  1.838(0.00) & 1.848(0.00) & 1.848(0.00)\\
\hline
$k_{\textrm{IC}} /$ &  EXP = 2.589 & AH & ETGA & VH\Tstrut \\
$[10^8 \textrm{s}^{-1}]$ &B3LYP     &  2.326(0.011) & 2.358(0.002) & 2.290(0.010)\\
                         &CAM-B3LYP &  4.674(0.017) & 4.700(0.012) & 4.627(0.016)\\
\hline
$\Phi_\text{QY}$ & EXP = 42.00 & AH & ETGA & VH \Tstrut \\
$\times 10^2$ &B3LYP     & 35.76(0.11) & 35.36(0.02) & 36.32(0.10)\\
              &CAM-B3LYP & 28.22(0.07) & 28.23(0.05) & 28.5409(0.07)\\
\hline
\end{tabular}
\end{ruledtabular}
\end{table}

 \begin{figure}
  \centering
  \includegraphics[width=.50\textwidth, trim=0 0.25cm 0 0.5cm, clip]{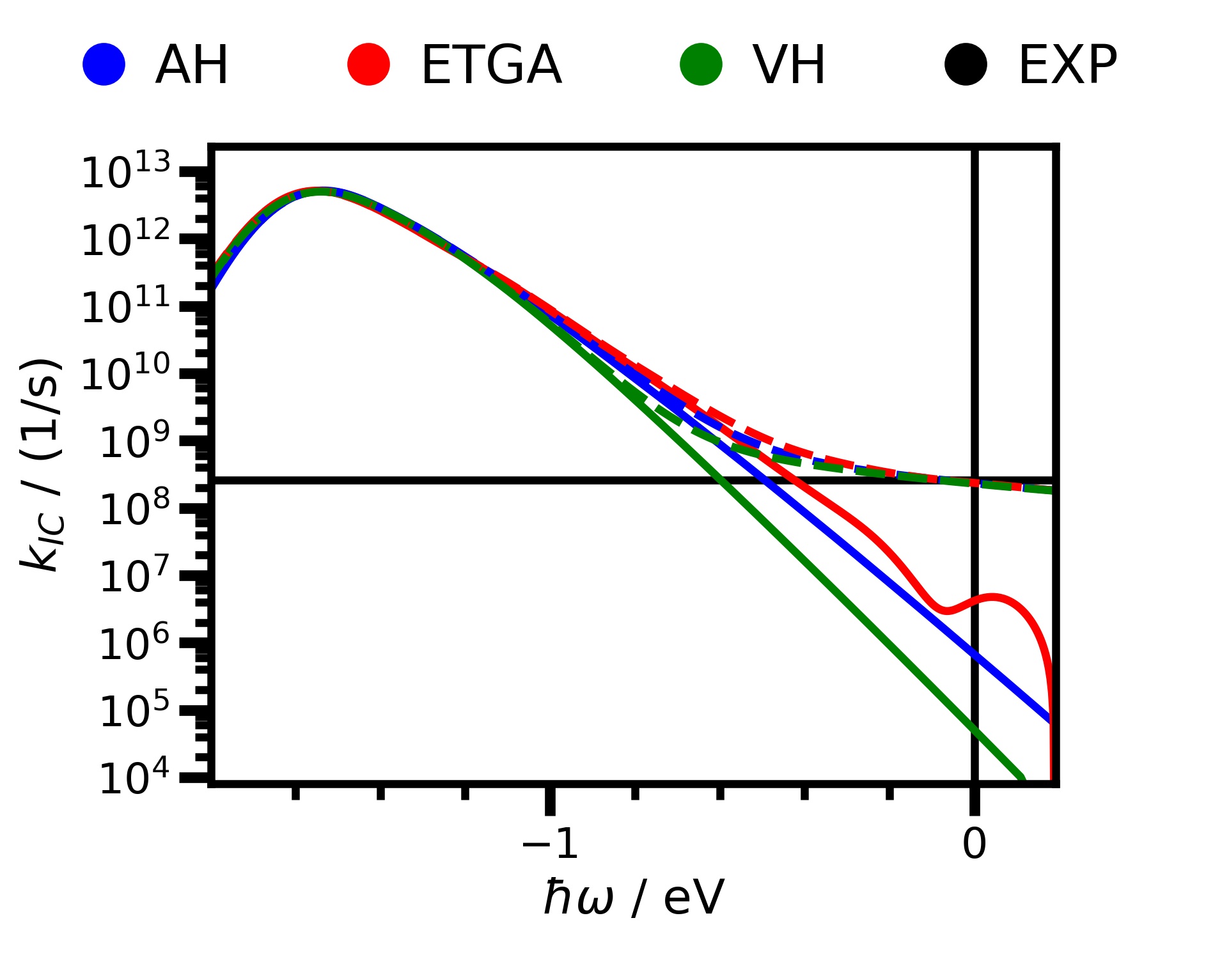}
\vspace{-20pt}
  \caption{Internal conversion spectrum of the $S_1\rightarrow S_0$ transition of the \textbf{squaraine dye}, calculated with the adiabatic harmonic model (AH), the extended thawed Gaussian approach (ETGA) and the vertical harmonic model (VH), using the B3LYP functional. The horizontal black line indicates the experimental rate, the vertical black line the point of energy conservation. The spectrum is based on a thermal correlation function (T=298.15 K) of 150.0 fs. The solid lines show the spectrum obtained using only a Gaussian line shape function with a HWHM value of 0.1 eV, the dashed lines show the spectrum using additionally a Lorentzian broadening function with a HWHM value of 0.001 eV.}
  \label{fig:sqm_kic_spectrum}
\end{figure}

\begin{figure}
  \centering
  \includegraphics[width=.50\textwidth, trim=0 0.25cm 0 0.3cm, clip]{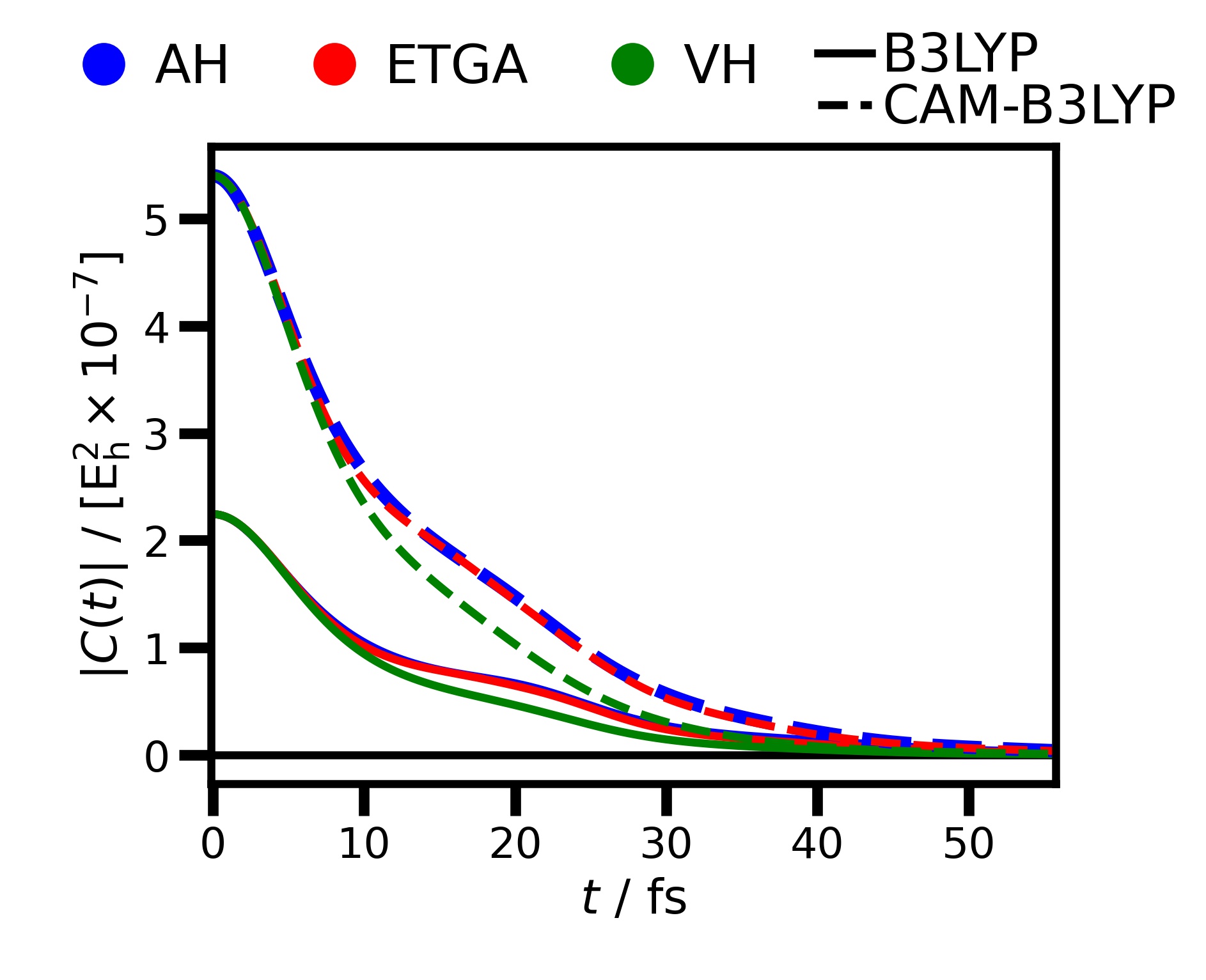}
\vspace{-20pt}
\setlength{\belowcaptionskip}{-6pt} 
  \caption{Magnitude of the thermal internal conversion correlation function (T=298.15 K) of the $S_1\rightarrow S_0$ transition of the \textbf{squaraine dye}, calculated with the adiabatic harmonic model (AH), the extended thawed Gaussian approach (ETGA) and the vertical harmonic model (VH), using B3LYP functional(solid lines) and CAM-B3LYP(dashed lines), without application of broadening functions.}
  \label{fig:sqm_kic_cor}
\end{figure}

Squaraine dyes are of interest due to their efficient emission at the low energy region of the visible spectrum, corresponding to red and near infrared.\cite{Mayerhoeffer2013} They are also used as building block supramolecular material design.\cite{Wuerthner2011,Gsaenger2014,Shen2020,Maeda2020}  With the energy gap law\cite{Englman1970} in mind one might have expected a low fluorescence quantum yield, as the internal conversion rate is often observed to rise exponentially with a decrease of the difference of the energy of the involved states. A reliable prediction of the internal conversion rate is thus particularly important in the search of efficient emitters with a small energy gap. The experimental quantum yield and fluorescence lifetime at room temperature in chloroform have been reported in the supplementary material of reference \onlinecite{Maeda2020}. This squaraine dye is well suited for theoretical study, as it is of moderate size with 64 atoms and contains only first and second row elements, DFT and TDDFT with B3LYP should thus be sufficiently accurate. Solvent effects on the energy of electronics states were accounted for with the polarizable continuum model for chloroform.\cite{Scalmani2010}

The harmonic approximation appears to be valid, indicated by the agreement of the adiabatic energy gap based on the optimized minimum and the energy gap based on the harmonic potential expanded at the vertical geometry, listed in Table \ref{tab:sqme_results}, together with the spontaneous emission rates, the internal conversion rates and quantum yields. The spontaneous emission rates vary little among the models, but the prediction of the internal conversion rate suffers again from a large dependency on the line shape function. A purely Gaussian broadening would lead to an underestimate of the internal conversion rate by at least an order of magnitude, shown in Fig. \ref{fig:sqm_kic_spectrum} for the B3LYP calculation with all models. The internal conversion spectrum using a pure Gaussian broadening shown in solid lines falls off too fast. But once combined with a Lorentzian, the theoretical values compare favorably with the experimental result. It is also noticeable that the difference between the internal conversion rates almost vanish with respect to the AH, VH and ETGA model. Variation due to the choice of the DFT functional remains; a difficult choice, as B3LYP performs well for the internal conversion rate while deviating in the case of the spontaneous emission rate. This situation is reversed in the case of CAM-B3LYP, which predicts emission rates that are close to the experiment, while the internal conversion rate is overestimated. The closest value to the experimental quantum yield of 42 \% is obtained with B3LYP and the VH model, but the differences between models are minor and the predicted values fall all within a range of 35.3--36.3 \%, for CAM-B3LYP all quantum yields fall within 28.2--28.5 \%.

The reported experimental emission spectrum consists of a single broad peak,\cite{Maeda2020} which is predicted by all three models besides a small energy shift of $\approx$0.1 eV. The spectra are given in Fig. S4 of the supplementary material.
The magnitude of the transition dipole moment vectors are $|\vec \mu^\mathrm{B3LYP}_{10}|^2=25.95$ au and $|\vec \mu^\mathrm{CAM-B3LYP}_{10}|^2=30.09$ au, the nonadiabatic coupling vector norms are $|\vec \tau^\mathrm{B3LYP}_{10}|^2=1.73$ au and $|\vec \tau^\mathrm{CAM-B3LYP}_{10}|^2=3.55$ au and their ratios are close to the ratios of the predicted rates, explaining the observed variation with respect to the functionals.  The difference is also clearly visible in Fig. \ref{fig:sqm_kic_cor}, showing the absolute value of the internal conversion correlation function for both functionals (solid line for B3LYP, dashed for CAM-B3LYP) and all models (color coded). The B3LYP and CAM-B3LYP calculations show the same progression besides the aforementioned constant scaling factor, due to the difference of nonadiabatic coupling vector magnitude. After an initial decay that is the same in all models, deviations start to appear around the 10 fs mark, and remain for a duration of roughly 30 fs, to converge and finally decay beyond 50 fs. A remarkable feature of these curves is the close agreement of the AH and ETGA model. This point is not of major importance for the rates of this specific dye, but it shows that ETGA method can be used to decide if the AH or VH is appropriate to use, and that the ETGA method combines the best of both. It  captures the initial decay properly due to the expansion of the potential around the vertical transition geometry in the beginning, while the time-dependent nature of the local harmonic approximation enables it to model also the dynamics once this region is left and a harmonic approximation around the equilibrium structure is preferable. 

% It can capture the initial decay as well as the VH model, due to the expansion of the potential around the vertical transition  geometry, while the time-dependent nature of the local harmonic approximation enables it model the dynamics also once this region is left and a harmonic approximation around the equilibrium structure is preferable. 

\section{Conclusion}

% Two global harmonic models, defined by their difference in the choice of the expansion point of the potential, the optimized geometry of the final electronic state of the transition for the adiabatic model versus the optimized geometry of the initial state for the vertical model, have been compared with the extended thawed  Gaussian approximation, an Ansatz based on a dynamic and implicitly time-dependent harmonic approximation by following a dynamic expansion point that is determined by the propagation of a classical trajectory on the true final potential. The application to 

Spontaneous emission and internal conversion rates have been calculated with the AH, VH and ETGA model across a test set for which reliable experimental data is available in the literature, comprised of formaldehyde, fluorobenzene, azulene and a squaraine dye. Significant differences are observed for the values of the internal conversion rate in the case of formaldehyde and azulene as seen in Tab. \ref{tab:formaldehyde_results} and Tab. \ref{tab:azulene_results}. The adiabatic and vertical model yield results that are more than an order of magnitude apart, but the ETGA model supports in both cases the vertical model. This agreement indicates that the ETGA and VH model are better choices for internal conversion rate calculations, which is also reflected in the comparison with the experimental values. The adiabatic harmonic model has been used in the calculation of intersystem crossing rates\cite{Banerjee2016} and could be useful when calculating nonradiative rates for systems with strong spin-orbit coupling. The energy gap between states involved in intersystem crossing can be low and could favor the adiabatic model, which is expected to be suitable in the treatment of transitions involving vibrational states of low vibrational quantum number.

Differences in the spontaneous emission rate are less significant and any of the three methods yields acceptable results when the absolute rate values are compared with the experiment. The ETGA predicts emission spectra that in good agreement with the experiment for all cases except formaldehyde. The main contribution to the error in the emission rate in fluorobenzene, azulene and the squaraine dye is thus most likely due to errors in the energies and in particular due to wrong magnitudes of the transition dipole moment. In our test systems rates were typically too small by a factor within a range of two to three. There is no comparable experiment that yields internal conversion spectra, but the fact that quantum yields are often obtained in good agreement to the experiment indicates that the nonadiabatic coupling elements are subject to similar errors.

The ETGA yields good results, but the difference to the VH results are small. Overall it appears reasonable to start out with the VH model due to its lower cost, but the ETGA method is a good choice to obtain clarity should a stalemate between the AH and VH method occur and can be used if the Hessian at the vertical geometry has negative eigenvalues, which is problematic for the VH model as the resulting harmonic approximation would be repulsive in the modes associated with the imaginary frequency.

The straightforward protocol that was adopted managed to predict quantum yields within one order of magnitude of the experimental value for formaldehyde and azulene, but failed for fluorobenzene and the squaraine dye. The internal conversion rate would have been underestimated in these two cases if the line shape function had not been adjusted; averaging over calculations with purely Gaussian line shape functions was insufficient. Results close to the experiment required the application of an additional Lorentzian broadening. This is problematic as this causes changes on multiple orders of magnitude - the variation due to the HWHM value of the Lorentzian is substantial and exceeds the influence of energies and coupling elements considerably in these cases. The internal conversion spectrum appears only to be robust and reliable in a limited interval close to the maximum of the internal conversion spectrum $k_{\textrm{IC}}(\omega_{\textrm{max}})$, where the value is not unduly sensitive to the choice of the line shape function. For fluorobenzene this interval seems to be given by values $k_{\textrm{IC}}(\omega)>10^{-4}k_{\textrm{IC}}(\omega_{\textrm{max}})$, for the squaraine dye we find $k_{\textrm{IC}}(\omega)>10^{-3}k_{\textrm{IC}}(\omega_{\textrm{max}})$. This issue was already discussed in great detail in Ref. \onlinecite{Humeniuk2020} for different global harmonic models and the reported findings are largely corroborated by our results.

This work proofed the viability of the ETGA method for the calculation of the internal conversion and spontaneous emission rates in molecules and revisited the adiabatic and vertical harmonic approximation but the line shape problem is not solved by including anharmonicity effects or improvements to the calculation of the internal conversion correlation function. Otherwise it would not have occurred in the squaraine dye, where all three models predict essentially the same, which happens when the harmonic approximation is independent of the expansion point, a clear indicator of a highly harmonic potential. The aspiration to predict internal conversion rates and radiative quantum yields is clearly in need of an additional theoretical model that fixes the choice of the line shape function and its width.

 % This work proofed the viability of the ETGA method for the calculation of the internal conversion and spontaneous emission rates in molecules and revisited the adiabatic and vertical harmonic approximation, but also shed light on remaining problem that the choice of the line shape function poses. The importance of this effect can not be understated, as it can change the results over multiple orders of magnitude and deserves more attention. There seems to be much room left for improvements in this regard.

\section*{Supplementary Material}
See the supplementary material for a comparison of experimental and calculated spontaneous emission spectra.

\begin{acknowledgments}
We gratefully acknowledge financial support by the Deutsche Forschungsgemeinschaft via grant MI1236/6-1.
\end{acknowledgments}

\section*{Author declarations}

\subsection*{Conflicts of Interest}

There are no conflicts to declare.

\subsection*{Author Contributions}

\noindent\textbf{Michael Wenzel}: Investigation (equal); Methodology (equal); Software (equal); Validation (equal); Visualization (equal); Writing – original draft (equal). \textbf{Roland~Mitric}: Conceptualization (equal); Funding acquisition (equal); Methodology (equal); Supervision (equal); Writing – review \& editing (equal).

\section*{Data Availability}
The data that support the findings of this study are available from the corresponding author upon reasonable request.

\section*{References}
\nocite{*}
\bibliography{aipsamp}% Produces the bibliography via BibTeX.

\end{document}

\maketitle

% \section{\label{sec:level1}First-level heading:\protect\\ The line
% break was forced \lowercase{via} \textbackslash\textbackslash}
\FloatBarrier

\newpage

 \begin{figure}[H]
  \centering
  \includegraphics[width=1.0\textwidth, trim=0 0.0cm 0 0.0cm, clip]{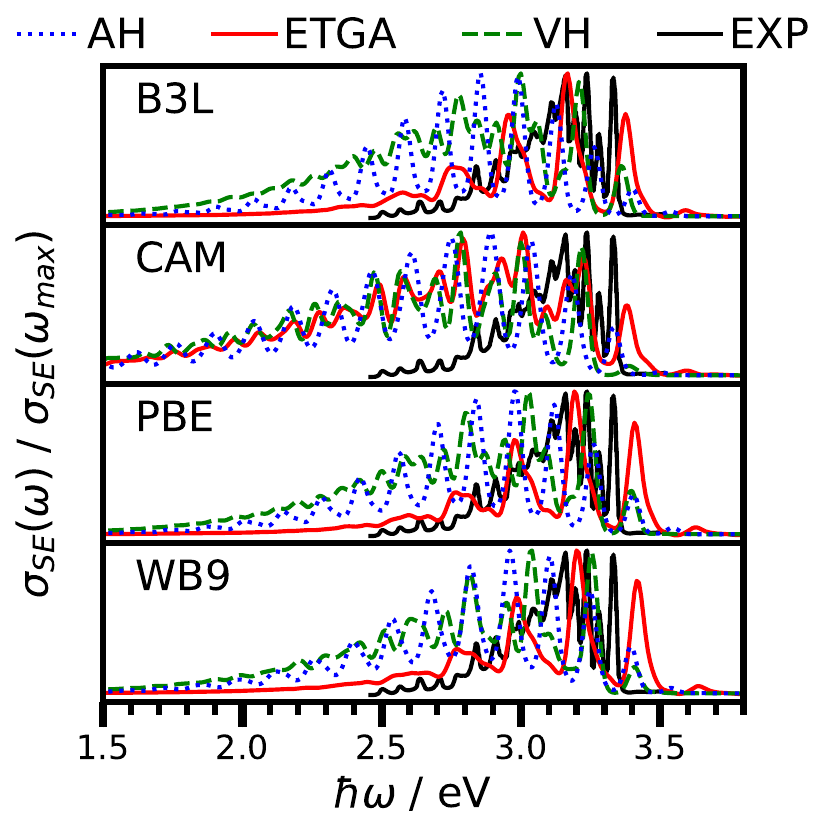}
  \caption{\textbf{Formaldehyde} S$_1$ spontaneous emission spectra. The experimental spectrum was taken from Ref. \cite{Smith2003} - Fig 1. The simulated spectra are based on DFT/TD-DFT calculations with the B3LYP (B3L), CAM-B3LYP (CAM), PBE0 (PBE) and $\omega$B97XD (WB9) functionals in combination with a 6-311G basis set, using the Adiabatich Harmonic model (AH), the Vertical Harmonic model (VH) and the Extended Thawed Gaussian Approximation (ETGA), with parameters as given in the main text, section III. A Gaussian lineshape function with a HWHM=0.025 eV was used for the broadening. The spectra are normalized by their highest peak. None of the spectra are shifted or corrected otherwise.} 
  \label{fig:formaldehyde_emission_spectra}
   
\end{figure}

\newpage

 \begin{figure}[H]
  \centering
  \includegraphics[width=1.0\textwidth, trim=0 0.0cm 0 0.0cm, clip]{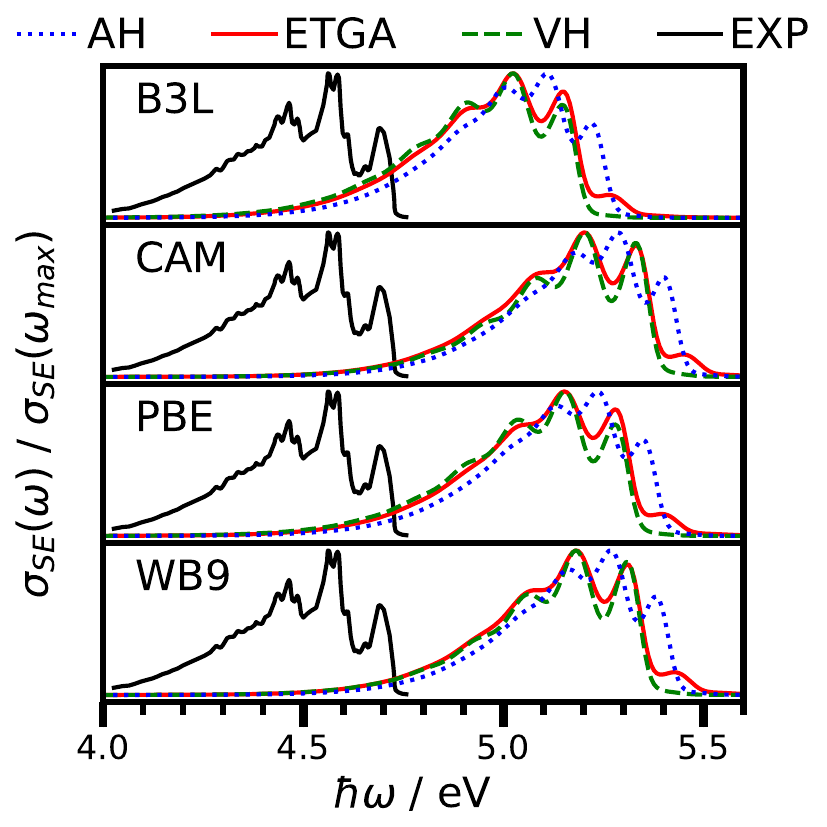}
  \caption{\textbf{Fluorobenzene} S$_1$ spontaneous emission spectra. The experimental spectrum was taken from Ref. \cite{Nakamura2003} - Figure 3, solid line. The simulated spectra are based on DFT/TD-DFT calculations with the B3LYP (B3L), CAM-B3LYP (CAM), PBE0 (PBE) and $\omega$B97XD (WB9) functionals in combination with a 6-311G basis set, using the Adiabatich Harmonic model (AH), the Vertical Harmonic model (VH) and the Extended Thawed Gaussian Approximation (ETGA), with parameters as specified in the main text, section III. A Gaussian lineshape function with a HWHM=0.025 eV in combination with a Lorentzian lineshape function with HWHM=0.001 eV were used for the broadening. The spectra are normalized by their highest peak. None of the spectra are shifted or corrected otherwise.}
  \label{fig:fluorobenzene_emission_spectra}
\end{figure}

\newpage

 \begin{figure}[H]
  \centering
  \includegraphics[width=1.0\textwidth, trim=0 0.0cm 0 0.0cm, clip]{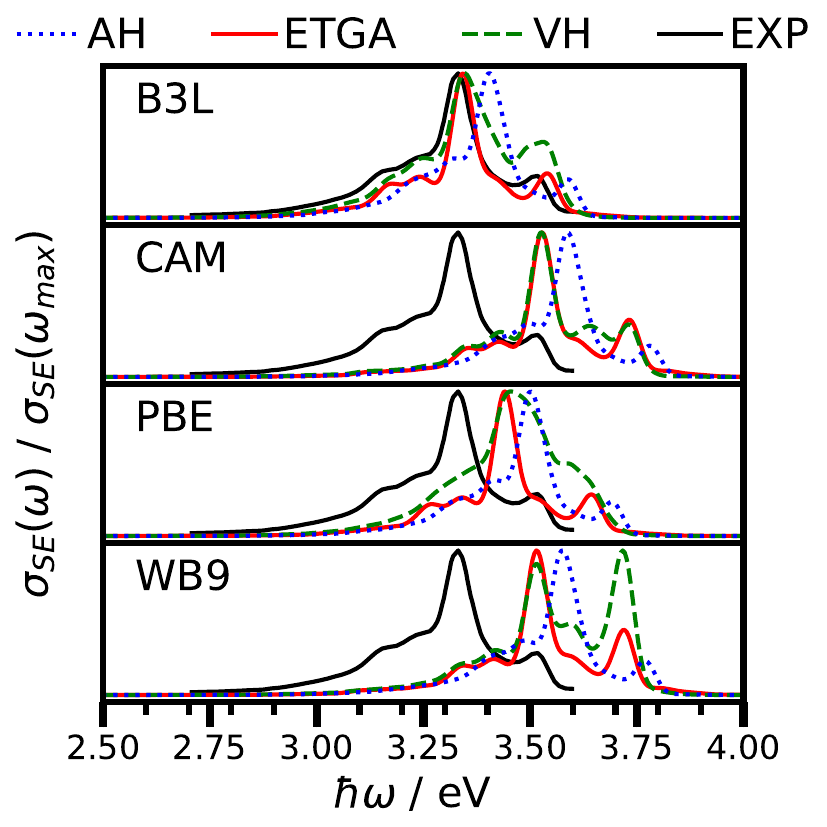}
  \caption{\textbf{Azulene} S$_2$ spontaneous emission spectra. The experimental spectrum was taken from Ref. \cite{Prlj2020}  - Figure 4 (c). The simulated spectra are based on DFT/TD-DFT calculations with the B3LYP (B3L), CAM-B3LYP (CAM), PBE0 (PBE) and $\omega$B97XD (WB9) functionals in combination with a 6-311G basis set, using the Adiabatich Harmonic model (AH), the Vertical Harmonic model (VH) and the Extended Thawed Gaussian Approximation (ETGA), with parameters as specified in the main text, section III. A Gaussian lineshape function with a HWHM=0.025 eV was used for the broadening. The spectra are normalized by their highest peak. None of the spectra are shifted or corrected otherwise.}
  \label{fig:squaraine_emission_spectra}
\end{figure}

\newpage

 \begin{figure}[H]
  \centering
  \includegraphics[width=1.0\textwidth, trim=0 0.0cm 0 0.0cm, clip]{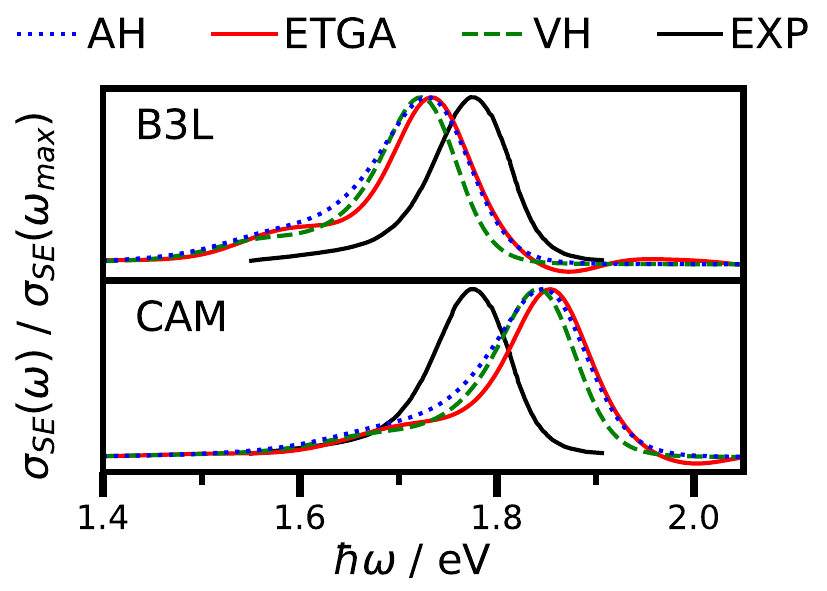}
  \caption{\textbf{Squaraine dye} S$_1$ spontaneous emission spectra. The experimental spectrum was taken from the \href{https://www.rsc.org/suppdata/d0/cc/d0cc04306k/d0cc04306k1.pdf}{Supplementary material} Supplementary material of Ref. \cite{Maeda2020} - Figure S4. The simulated spectra are based on DFT/TD-DFT calculations with the B3LYP (B3L) and CAM-B3LYP (CAM) functionals in combination with a 6-311G basis set, using the Adiabatich Harmonic model (AH), the Vertical Harmonic model (VH) and the Extended Thawed Gaussian Approximation (ETGA), with parameters as specified in the main text, section III. A Gaussian lineshape function with a HWHM=0.025 eV in combination with a Lorentzian lineshape function with HWHM=0.001 eV were used for the broadening. The spectra are normalized by their highest peak. None of the spectra are shifted or corrected otherwise.}
  \label{fig:squaraine_emission_spectra}
\end{figure}

\hfill

\FloatBarrier
\printbibliography